\documentclass[oupdraft]{bio}
\RequirePackage{amsthm,multirow,amsmath,fancybox,amssymb}
\RequirePackage{graphicx}
\RequirePackage{lastpage}
\usepackage{url,mathrsfs}
\usepackage{setspace}
\usepackage{textcomp,color}
\usepackage{rotating}
\usepackage{natbib}
\def\half{\frac{1}{2}}
\def\nhalf{n^{\half}}
\def\nnhalf{n^{-\half}}
\def\bZ{\mathbf{Z}}
\def\Dscr{\mathscr{D}}

\def\bD{\mathbf{D}}
\def\bG{\mathbf{G}}

\def\transpose{{\sf \scriptscriptstyle{T}}}
\def\trans{^{\transpose}}
\def\bbeta{\boldsymbol{\beta}}
\def\Asc{\mathcal{A}}
\def\Aschat{\widehat{\mathcal{A}}}
\def\Ssc{\mathcal{S}}
\def\Sschat{\widehat{\mathcal{S}}}
\def\subfull{_{\sf \scriptscriptstyle{full}}}
\def\bbetahat{\widehat{\bbeta}}
\def\betahat{\widehat{\beta}}
\def\betatilde{\widetilde{\beta}}
\def\bbetatilde{\widetilde{\bbeta}}
\def\bUhat{\widehat{\bU}}
\def\bUtilde{\widetilde{\bU}}
\def\ellhat{\widehat{\ell}}
\def\Abb{\mathbb{A}}
\def\Abbhat{\widehat{\Abb}}

\def\bU{\mathbf{U}}

\def\ba{\mathbf{a}}
\def\bX{\mathbf{X}}
\def\bY{\mathbf{Y}}

\def\sDAC{{\sf \scriptscriptstyle{DAC}}}
\def\subDAC{_{\sf \scriptscriptstyle{DAC}}}

\def\subind{_{\sf \scriptscriptstyle{ind}}}
\def\subindt{_{\sf \scriptscriptstyle{ind},0}}
\def\subdep{_{\sf \scriptscriptstyle{dep}}}
\def\subdept{_{\sf \scriptscriptstyle{dep},0}}
\def\subDACj{_{{\sf \scriptscriptstyle{DAC}},\jmath}}
\def\DAC{\mbox{DAC}}
\def\subklin{_{k,\sf \scriptscriptstyle{lin}}}
\def\sublin{_{\sf \scriptscriptstyle{lin}}}
\def\subfulllin{_{\sf \scriptscriptstyle{full,lin}}}
\def\subMV{_{\sf\scriptscriptstyle MV}}
\def\supA{^{\scriptscriptstyle{\Asc}}}
\def\argmindum{\mathop{\mbox{argmin}}}
\def\argmin#1{\argmindum_{#1}}
\def\argmaxdum{\mathop{\mbox{argmax}}}
\def\argmax#1{\argmaxdum_{#1}}
\def\ninv{n^{-1}}

\def\supiota{^{[\iota]}}
\def\sumkK{\sum_{k=1}^K}
\def\Kinv{K^{-1}}
\def\Isc{\mathcal{I}}
\definecolor{darkred}{RGB}{150,50,50}
\definecolor{brown}{RGB}{250,100,100}
\definecolor{green}{RGB}{000,150,100}
\definecolor{purple}{RGB}{250,000,180}

\begin{document}
% Title of paper
\title{\vspace{-5ex} \usefont{\encodingdefault}{\rmdefault}{b}{n}%
	\fontsize{16}{20}%
	\selectfont A Fast Divide-and-Conquer Sparse Cox Regression}

% List of authors, with corresponding author marked by asterisk
\author{Yan Wang,$^{1,2}$ Nathan Palmer,$^3$ Qian Di,$^1$ Joel Schwartz,$^1$  Isaac Kohane,$^3$ Tianxi Cai$^{2,3,*}$\\
\textit{1. Department of Environmental Health, Harvard T.H. Chan School of Public Health}\\
\textit{2. Department of Biostatistics, Harvard T.H. Chan School of Public Health}\\
\textit{3. Department of Biomedical Informatics, Harvard Medical School}\\
* Correspondence to \textit{tcai@hsph.harvard.edu}}
% Running headers of paper:
\markboth%
% First field is the short list of authors
{Y. Wang and others}
% Second field is the short title of the paper
{A Fast Divide-and-Conquer Sparse Cox Regression}

\maketitle
%\newpage
% Add a footnote for the corresponding author if one has been
% identified in the author list

\begin{abstract}
{We propose a computationally and statistically efficient divide-and-conquer (DAC) algorithm to fit sparse Cox regression to massive datasets where the sample size $n_0$ is exceedingly large and the covariate dimension $p$ is not small but $n_0\gg p$. The proposed algorithm achieves computational efficiency through a one-step linear approximation followed by a least square approximation to the partial likelihood (PL). These sequences of linearization enable us to maximize the PL with only a small subset and perform penalized estimation via a fast approximation to the PL. The algorithm is applicable for the analysis of both time-independent and time-dependent survival data. Simulations suggest that the proposed DAC algorithm substantially outperforms the full sample-based estimators and the existing DAC algorithm with respect to the computational speed, while it achieves similar statistical efficiency as the full sample-based estimators. The proposed algorithm was applied to an extraordinarily large time-independent survival dataset and an extraordinarily large time-dependent survival dataset for the prediction of heart failure-specific readmission within 30 days among Medicare heart failure patients.}{Divide-and-conquer; shrinkage estimation; variable selection; Cox proportional hazards model; least square approximation.}
\end{abstract}

\newpage

\section{Introduction}
\label{sec1}
Large datasets derived from health insurance claims and electronic health records are becoming increasingly available for health care and medical research. These datasets serve as valuable sources for the development of risk prediction models, which are the key components of precision medicine. Fitting risk prediction models to a dataset with a massive sample size ($n_0$), however, is computationally challenging, especially when the number of candidate predictors ($p$) is also large and yet only a small subset of the predictors are informative. In such a setting, it is highly desirable to fit a sparse regression model to simultaneously remove non-informative predictors and estimate the effects of the informative predictors. When the outcome of interest is time-to-event and is subject to censoring, one may obtain a sparse risk prediction model by fitting a regularized Cox proportional hazards model \citep{cox1972regression} with penalty functions such as the adaptive least absolute shrinkage and selection operator (LASSO) penalty \citep{Zhang2007}.

When $n_0$ is extraordinarily large, directly fitting an adaptive LASSO penalized Cox model to such a dataset is not computationally feasible. To overcome the computational difficulty, one may employ the divide-and-conquer (DAC) strategy, which typically divides the full sample into subsets, solves the optimization problem using each subset, and combines the subset-specific estimates into a combined estimate. Various DAC algorithms have been proposed to fit penalized regression models. For example, \cite{Chen2014} proposed a DAC algorithm to fit penalized generalized linear models (GLM). The algorithm obtains a sparse GLM estimate for each subset and then combines subset-specific estimates by majority voting and averaging. \cite{Tang2016} proposed an alternative DAC algorithm to fit GLM with an extremely large $n_0$ and a large $p$ by combining de-biased LASSO estimates from each subset. While both algorithms are effective in reducing the computation burden compared to fitting a penalized regression model to the full data, they remain computationally intensive as $K$ penalized estimation procedures will be required. In addition, the DAC strategy has not been extended to the survival data analysis.

In this paper, we propose a novel DAC algorithm using sequences of linearization, denoted by $\DAC\sublin$, to fit adaptive LASSO penalized Cox proportional hazards models, which can further reduce the computation burden compared to the existing DAC algorithms. $\DAC\sublin$ starts with obtaining an estimator that maximizes the partial likelihood (PL) of a subset of the full data, which is then updated using all subsets via one-step approximations. The updated estimator serves as a $\sqrt{n_0}$-consistent initial estimator for the adaptive LASSO problem and approximates the full sample-based maximum PL estimator. Subsequently, we obtain the final adaptive LASSO estimator based on an objective function applying the least square approximation (LSA) to the PL as in \cite{Wang2007}. The LSA allows us to fit the adaptive LASSO using a pseudo likelihood based on a sample of size $p$. The penalized regression is only fit once in the proposed $\DAC\sublin$ algorithm and the improvement in computation cost is substantial if $n_0\gg p$. Our proposed $\DAC\sublin$ algorithm can also accommodate time-dependent covariates.

The rest of the paper is organized as follows. We detail the $\DAC\sublin$ algorithm in section \ref{sec-method}. In section \ref{sec-sim}, we present simulation results demonstrating the superiority of $\DAC\sublin$ compared to the existing methods when covariates are time-independent and when some covariates are time-dependent. In section \ref{sec-app}, we employ the $\DAC\sublin$ algorithm to develop risk prediction models for 30-day readmission after an index heart failure hospitalization with data from over 10 million Medicare patients by fitting regularized Cox models with (i) $p=540$ time-independent covariates and (ii) $p\subind=575$ time-independent covariates and $p\subdep=5$ time-dependent environmental covariates. We conclude with some discussions in section \ref{sec-discuss}.

\section{Methods}
\label{sec-method}
\subsection{Notation and Settings}
Let $T$ denote the survival time and $\bZ(\cdot)$ denote the $p\times 1$ vector of bounded and potentially time-dependent covariates. Due to censoring, for $T$, we only observe $(X, \Delta)$, where $X = T \wedge C$, $\Delta = I(T \le C)$, and $C$ is the censoring time assumed to be independent of $T$ given $\bZ(\cdot)$. Suppose the data for analysis consist of $n_0$ subjects with independent realizations of $\bD = (X, \Delta, \bZ(\cdot)\trans)\trans$, denoted by $\Dscr\subfull = \{\bD_i = (X_i,\Delta_i,\bZ_i(\cdot)\trans)\trans, i = 1, ..., n_0\}$, where we assume that $n_0 \gg p$. 

We denote the index set for the full data by $\Omega\subfull = \{1, ..., n_0\}$. For all DAC algorithms discussed in this paper, we randomly partition $\Dscr\subfull$ into $K$ subsets with the $k$-th subset denoted by $\Dscr_k = \{\bD_i, i \in \Omega_k\}$. Without loss of generality, we assume that $n = n_0/K$ is an integer and that the index set for the subset $k$ is $\Omega_k = \{(k-1)n+1, ..., kn\}$. For any index set $\Omega$, we denote the size of $\Omega$ by $n_{\Omega}$ with $n_{\Omega}=n_0$ if $\Omega=\Omega\subfull$ and $n_{\Omega}=n$ if $\Omega=\Omega_k$. Throughout we assume that $K = o\left(\nhalf_0\right)$ such that $\ninv = o\left(\nnhalf_0\right)$ and $n\gg p$. 

To develop a risk prediction model for $T$ based on $\bZ(\cdot)$, we consider the Cox model,
\begin{equation}{\label{Cox}}
\lambda(t|\bZ(t)) = \lambda_0(t)\exp\left( \beta_0\trans \bZ(t)\right),
\end{equation}
where $\lambda(t|\bZ(t))$ is the conditional hazard function and $\lambda_0(t)$ is the baseline hazard function. Our goal is to develop a computationally and statistically efficient procedure to estimate $\bbeta_0$ using data in $\Dscr\subfull$ under
the assumption that $\bbeta_0$ is sparse with the size of the active set $\Asc = \{\jmath: \beta_{0\jmath} \ne 0\}$ much smaller than $p$. 

When $n_0$ is not extraordinarily large, we may obtain an efficient estimate, denoted by $\bbetahat\subfull$, based on the adaptive LASSO penalized PL likelihood estimator as proposed in \cite{Zhang2007}. Specifically,
\begin{equation}{\label{penLP}}
\bbetahat\subfull =  \argmax{\bbeta} \left\{\ellhat_{\Omega\subfull}(\bbeta)-\lambda_{\Omega\subfull}\sum_{\iota=1}^p\frac{|\beta_\iota |}{|\widetilde{\beta}_{\iota,\text{init}}|^\gamma} \right\}
\end{equation}
where for any index set $\Omega$, 
\begin{equation}{\label{likelihood}}
\ellhat_{\Omega}(\bbeta) =  \ninv_{\Omega} \sum_{i \in \Omega}\ell_{i}(\bbeta) , \quad \ell_i(\bbeta) = \Delta_i\left[\bbeta\trans \bZ_{i}(X_i)-\log\left\{\sum_{i' \in \Omega} I(X_{i'} \ge X_i)e^{\bbeta\trans \bZ_{i'}(X_{i'})}\right\}\right],
\end{equation}
$\bbetatilde_{\text{init}}=(\widetilde{\beta}_{1,\text{init}},\cdots,\widetilde{\beta}_{p,\text{init}})'$ is an initial $\sqrt{n_0}$-consistent estimator of model \eqref{Cox}, $\lambda_{\Omega\subfull}\ge 0$ is a tuning parameter, and $\gamma>0$. A simple choice of $\bbetatilde_{\text{init}}$ is $\bbetatilde_{\Omega\subfull}$, where for any set $\Omega$, 
$$\bbetatilde_{\Omega} = \argmax{\bbeta}{\ellhat_{\Omega}(\bbeta)}.$$ 
Following the arguments given in \cite{Zhang2007}, when $\nhalf_0\lambda_{\Omega\subfull} \to 0$ and $n_0^{(1+\gamma)/2} \lambda_{\Omega\subfull} \to \infty$, we can show that $\bbetahat\subfull$ achieves the variable selection consistency, i.e. the estimated active set $\Aschat{\subfull}=\{\jmath:\widehat\beta{\subfull}_{,\jmath}\ne 0 \}$ satisfies $P(\Aschat{\subfull} = \Asc) \to 1$ and that the oracle property holds, i.e.
$$
\nhalf_0(\bbetahat\subfull\supA - \bbeta_0) = \Abbhat_{\Omega\subfull}\supA(\bbeta_0)^{-1}\nhalf_0  \bUhat_{\Omega\subfull}\supA(\bbeta_0)  + o_p(1) = \Abb\supA(\bbeta_0)^{-1}\nhalf_0  \bUhat_{\Omega\subfull}\supA(\bbeta_0)  + o_p(1)\xrightarrow{D}\mathcal{N}\left(\textbf{0},\Abb\supA(\bbeta_0)^{-1}\right), 
$$ 
where for any set $\Asc$, $\bG^{\Asc} = \{G_l, l \in \Asc\}$ if $\bG$ is a vector and $\bG^{\Asc} = [W_{ll'}]_{l \in \Asc,l' \in \Asc}$ if $\bG$ is a matrix, 
$$
\Abb(\bbeta) = \int \frac{\Ssc_2(t,\bbeta)\Ssc_0(t,\bbeta) - \Ssc_1(t,\bbeta)^{\otimes 2}}{\Ssc_0(t,\bbeta)^2} dE\{N_i(t)\},  \quad \Abbhat_{\Omega}(\bbeta)= -\frac{\partial^2 \ellhat_{\Omega}(\bbeta)}{\partial \bbeta\partial\bbeta\trans}
$$ 
$\bUhat_{\Omega}(\bbeta) = \ninv_{\Omega}\sum_{i \in \Omega} \bUhat_{i, \Omega}(\bbeta)$, $\Ssc_r(t, \bbeta) = E\{\bZ_i(t)^{\otimes r} I(X_i \ge t)\}$, $\Sschat_{r,\Omega}(t,\bbeta) = n_{\Omega}^{-1}\sum_{i \in \Omega} \bZ_i(t)^{\otimes r} I(X_i \ge t)$, 
$\bUhat_{i, \Omega}(\bbeta) = \int \{\bZ_i(t) - \Sschat_{1,\Omega}(t,\bbeta)/\Sschat_{0,\Omega}(t,\bbeta) \}d M_i(t,\bbeta)$, $N_i(t)=I(T_i\le t)\Delta_i$, 
$M_i(t, \bbeta) = N_i(t) - \int_0^t I(X_i \ge u) e^{\bbeta\trans\bZ_i(u)}\lambda_0(u)du$, $\ba^{\otimes 0} = 1$, $\ba^{\otimes 1} = \ba$ and $\ba^{\otimes 2} = \ba\ba\trans$ for any vector $\ba$. 

When $n_0$ is not too large, multiple algorithms are available to solve \eqref{penLP} with time-independent covariates, including a gradient descent algorithm (\citealp{Simon2011}), a least angle regression (LARS)-like algorithm (\citealp{Park2007}), a combination gradient descent-Newton Raphson method (\citealp{Goeman2010}), and a modified shooting algorithm (\citealp{Zhang2007}). Unfortunately, when $n_0$ is extraordinarily large, none of the existing algorithms for \eqref{penLP} will be computationally feasible. While these algorithms may be extended to fit sparse Cox models with time-dependent covariates, the computation is even more demanding since each subject may contribute multiple observations in the fitting.

\subsection{The $\DAC\sublin$ Algorithm}
\label{sub2.1}
The goal of this paper is to develop an estimator that achieves the same asymptotic efficiency as $\bbetahat\subfull$ but can be computed very efficiently.

Our proposed algorithm, $\DAC\sublin$, for attaining such a property is motivated by the LSA proposed in \cite{Wang2007}, with the LSA applied to the full sample-based PL. Specifically, it is not difficult to show that $\bbetahat\subfull$ is asymptotically equivalent to $\bbetahat\subfulllin$, where
$$
\bbetahat\subfulllin = \argmin{\bbeta}\half (\bbetatilde_{\Omega\subfull} - \bbeta)\trans \Abbhat_{\Omega\subfull}(\bbetatilde_{\Omega\subfull}) (\bbetatilde_{\Omega\subfull} - \bbeta) +  \lambda_{n_0}\sum_{\iota=1}^p\frac{|\beta_\iota |}{|\widetilde{\beta}_{\iota,\Omega\subfull}|^\gamma}  
$$
That is, $\bbetahat\subfulllin$ will also achieve the variable selection consistency as $\bbetahat\subfull$ and $\bbetahat\subfulllin\supA$ has the same limiting distribution as that of $\bbetahat\subfull\supA$. This suggests that an estimator can recover the distribution of $\bbetahat\subfull$ if we can construct an accurate DAC approximations to $\bbetatilde_{\Omega\subfull}$ and $\Abbhat_{\Omega\subfull}(\bbetatilde_{\Omega\subfull})$. To this end, we propose a linearization-based DAC estimator, denoted by $\bbetahat\subDAC$, which requires three main steps: (i) obtaining an estimator for the unpenalized problem $\bbetatilde\subDAC^{[0]}$ based on a subset, say $\Dscr_1$; (ii) obtaining updated estimators for the unpenalized problem through one-step approximations using all $K$ subsets; and (iii) constructing an adaptive LASSO penalized estimator based on LSA. The procedure also brings a $\Abbhat\subDAC(\bbetatilde\subDAC)$ that well approximates $\Abbhat_{\Omega\subfull}(\bbetatilde_{\Omega\subfull})$.

Specifically, in step (i), we use subset $\Dscr_1$ to obtain a standard maximum PL estimator, 
\begin{flalign*}
\text{\underline{step (i)}}&& \bbetatilde\subDAC^{[0]} \equiv \bbetatilde_{\Omega_1} = \argmax{\bbeta}{\ellhat_{\Omega_1}(\bbeta)}.&&\phantom{\text{(i)}}
\end{flalign*}
In step (ii), we obtain a DAC one-step approximation to $\bbetatilde_{\Omega\subfull}$,
\begin{flalign*}
\text{\underline{step (ii)}} &&\mbox{for $\iota = 1, ..., \Isc$,} \hspace{.2in} \bbetatilde\subDAC \supiota  = K^{-1}\sum_{k=1}^K \bbetatilde_{\Omega\subklin} (\bbetatilde\subDAC^{[\iota -1]})&&\phantom{\text{(ii)}}
\end{flalign*}
where
\begin{equation}
\bbetatilde_{\Omega\subklin}(\bbeta) =\bbeta +  \Abbhat\subDAC(\bbeta )^{-1}\bUhat_{\Omega_k}(\bbeta) \quad
\mbox{and}\quad \Abbhat\subDAC(\bbeta) = \Kinv\sumkK \Abbhat_{\Omega_k}(\bbeta) .
\label{def-1step-bk}
\end{equation}
Let $\bbetatilde\subDAC = \bbetatilde\subDAC^{[\Isc]}$ be our DAC approximation to $\bbetatilde_{\Omega\subfull}$. In practice, we find that it suffices to let $\Isc = 2$. 
Finally, we apply the LSA to the PL and approximate $\bbetahat\subfull$ using $\bbetahat\subDAC$, where 
\begin{flalign*}
\text{\underline{step (iii)}}&& 
\bbetahat\subDAC=\argmin{\bbeta}{\left\{\half (\bbetatilde\subDAC - \bbeta)\trans\Abbhat\subDAC(\bbetatilde\subDAC)(\bbetatilde\subDAC - \bbeta) 
+ \lambda_{\Omega{\subfull}}\sum_{\jmath = 1}^p \frac{|\beta_{\jmath}|}{|\betatilde\subDACj|^{\gamma}}\right\}}.&&\phantom{\text{(i)}}
\end{flalign*}
The optimization problem in step (iii) is equivalent to 
\begin{equation}{\label{PEN}}
\bbetahat\subDAC=\argmin{\bbeta}{\left\{\half (\widetilde{\bY}_0(\bbetatilde\subDAC) - \widetilde{\bX}_0(\bbetatilde\subDAC)\bbeta)\trans(\widetilde{\bY}_0(\bbetatilde\subDAC) - \widetilde{\bX}_0(\bbetatilde\subDAC)\bbeta) 
	+ \lambda_{\Omega{\subfull}}\sum_{\jmath = 1}^p \frac{|\beta_{\jmath}|}{|\betatilde\subDACj|^{\gamma}}\right\}}
\end{equation}
where $\tilde{\bY}_0(\bbetatilde\subDAC)=\Abbhat\subDAC(\bbetatilde\subDAC)^{\half}\bbetatilde\subDAC$ is a $p\times 1$ vector and $\tilde{\bX}_0(\bbetatilde\subDAC)=\Abbhat\subDAC(\bbetatilde\subDAC)^{\half}$ is a $p\times p$ matrix. The linearization allows us to solve the penalized regression step using a pseudo likelihood based on a sample of size $p$. The computation cost of this step compared to solving \eqref{penLP} reduces substantially when $n_0\gg p$. In the Appendix, we show that
$\nhalf_0(\bbetatilde\subDAC - \bbetatilde_{\Omega\subfull}) = o_p(1)$. It then follows from the similar arguments given in \cite{Wang2007}
that if  $n_0^{\half}\lambda_{n_0}\rightarrow0$, $n_0^{(1+\gamma)/2}\lambda_{n_0}\rightarrow\infty$, the estimated active set using $\DAC\sublin$ $\widehat\Asc\subDAC$ achieves the variable selection consistency, i.e. $P(\widehat\Asc\subDAC=\Asc)\rightarrow1$ and the oracle property holds, i.e. $\bbetahat\subDAC^{\Asc}$ and $\bbetahat\subfull^{\Asc}$ have the same limiting distribution. 

\subsection{Tuning and Standard Error Calculation}\label{sec-tuning}

The tuning parameter $\lambda_{\Omega{\subfull}}$ is chosen by minimizing the Bayesian information criteria (BIC) of the fitted model. \cite{Volinsky2000} showed that the exact Bayes factor can be better approximated for the Cox model if the number of uncensored cases, $d_{0} = \sum_{i \in \Omega\subfull} \Delta_i$, 
is used to penalize the degrees of freedom in the Bayesian Information Criteria (BIC). Specifically, for any given tuning parameter
$\lambda_{\Omega\subfull}$ with its corresponding estimate of $\bbeta$, $\bbetahat_{\lambda_{\Omega\subfull}}$, the BIC suggested by \cite{Volinsky2000} is defined as
\begin{equation}
\text{BIC}_{V,\lambda_{\Omega\subfull}}=-2 \sum_{i \in \Omega\subfull} \ell_i(\bbetahat_{\lambda_{\Omega\subfull}})+ (\log d_{0}) df_{\lambda_{\Omega\subfull}},
\end{equation}
where $df_{\lambda_{\Omega\subfull}}=\sum_{\jmath = 1}^p I(\betahat_{\lambda_{\Omega\subfull}, \jmath} \ne 0)$. 
With the LSA, we may further approximate $\text{BIC}_{V,\lambda_{\Omega\subfull}}$ by
\begin{equation}
\text{BIC}_{V_L,\lambda_{\Omega\subfull}}= n_{0} (\bbetahat\subDAC-\bbetahat_{\lambda_{\Omega\subfull}})\trans \Abbhat\subDAC(\bbetatilde\subDAC)(\bbetahat\subDAC-\bbetahat_{\lambda_{\Omega\subfull}})+ (\log d_{0}) df_{\lambda_{\Omega\subfull}}.
\end{equation}

For the estimation of $\bbetahat\subDAC$, we chose a $\lambda_{\Omega\subfull}$ such that $\text{BIC}_{V_L,\lambda_{\Omega\subfull}}$ is minimized. The oracle property is expected to hold in the setting where $n_0 \gg p$ and $n_0$ is extraordinarily large. We may thus estimate the variance-covariance matrix for $\nhalf_0(\bbetahat\supA\subDAC-\bbeta_0\supA)$ using $\Abbhat\supA(\bbetatilde\subDAC)^{-1}$. For $\jmath \in \Asc$, a $(1-\alpha)\times 100\%$ confidence interval for $\beta_{0\jmath}$ can be calculated accordingly.

\def\Vbb{\mathbb{V}}

\section{Simulations} \label{sec-sim}
\subsection{Simulation Settings}
We performed two sets of simulations to evaluate the performance of $\bbetahat\subDAC$ for the fitting of sparse Cox models, one with only time-independent covariates and the other with time-dependent covariates. For both scenarios, we let $n_0 = 1,000,000$ and $K=100$. We consider the number of iterations $\Isc=1,2,$ and $3$ to examine the impact of $\Isc$ on the proposed estimator.

\subsubsection{Time-independent covariates}  We conducted extensive simulations to evaluate the performance of the proposed estimator $\bbetahat\subDAC$ relative to (a) the performance of the full sample-based adaptive LASSO estimator for the Cox model $\bbetahat\subfull$ and (b) a majority voting-based DAC method for the Cox model, denoted by $\bbetahat\subMV$ also with $K=100$, penalized by a minimax concave penalty (MCP), which extends the majority voting-based DAC scheme for GLM proposed by \cite{Chen2014}. The reason of choosing $\bbetahat\subMV$ as a comparison is that there is no other DAC method available for the Cox model and only \cite{Chen2014} considered a similar majority voting-based DAC method for the penalized GLM with non-adaptive penalties. We set a \textit{priori} that $\bbetahat\subMV$ sets the estimate of a coefficient at zero, if at least 50\% of the subset-specific estimates have a zero estimate for that coefficient. In addition, we compared the performance of the DAC estimator $\bbetatilde\subDAC$ relative to the full sample maximum PL estimator $\bbetatilde_{\Omega\subfull}$.

For the penalized procedures, we selected the tuning parameter based on the BIC criterion discussed in section \ref{sec-tuning}. The adaptive LASSO procedures were fit using the \texttt{glmnet} function in R with $\gamma=1$; the MCP procedures were fit using the \texttt{ncvsurv} function in R. 

For the covariates, we considered $p=50$ and $p=200$. We generated  $\bZ$ from a multivariate normal distribution with mean $\textbf{0}_p\trans$ and variance-covariance matrix $\Vbb=[I(l=l')+vI(l \ne l')]_{l=1,...,p}^{l'=1,...,p}$, where $\ba_q$ denotes a $q\times 1$ vector with all elements being $a$ and we considered $v = 0.2$, $0.5$, and $0.8$ to represent weak, moderate, and strong correlations among the covariates. For a given $\bZ_i,i=1,\cdots,n_0$, we generated $T_i$ from a Weibull distribution with a shape parameter of 2 and a scale parameter of $\{0.5\exp(\bbeta_0\trans \bZ_i)\}^{-0.5}$, where we considered three choices of $\bbeta_0$ to reflect different degrees of sparsity and signal strength: 
\begin{align*}
\bbeta_0^{\sf(I)} & =(\mathbf{0.8}_{3}\trans, \mathbf{0.4}_{3}\trans,\mathbf{0.2}_{3}\trans,\mathbf{0}_{p-9}\trans)\trans, \\
\bbeta_0^{\sf(II)} & =(\mathbf{0.4}_{4}\trans, \mathbf{0.2}_{4}\trans,\mathbf{0.1}_{4}\trans,\mathbf{0.05}_{4}\trans,\mathbf{0}_{p-9}\trans)\trans, \quad \mbox{and} \\
\bbeta_0^{\sf(III)}&=(1,0.5,\mathbf{0.2}_2\trans,\mathbf{0.1}_2\trans,\mathbf{0.05}_{2}\trans,\mathbf{0.035}_3\trans,\mathbf{0}_{p-11})\trans.
\end{align*} 
For censoring, we generated $C$ from an exponential distribution with a rate parameter of $\exp(0.5)$, resulting in $68\% \sim 76\%$ of censoring across different configurations.

\subsubsection{Time-dependent covariates} We also conducted simulations for the settings where time-dependent covariates are present to evaluate the performance of $\bbetahat\subDAC$. Since neither \texttt{glmnet} nor \texttt{ncvsurv} allows time-dependent survival data, we used $\bbetahat\subfulllin$ as a benchmark to compare $\bbetahat\subDAC$ with. In addition, we compared the performance of $\bbetatilde\subDAC$ relative to  $\bbetatilde_{\Omega\subfull}$.

We considered $p=100$ consisting of $p\subind=50$ time-independent covariates and $p\subdep=50$ time-dependent covariates. The simulation of the survival data with time-dependent covariates extended the simulation scheme of \cite{Austin2012} from dichotomous time-dependent covariates to continuous time-dependent covariates. We considered four time intervals $R_1=[0,1)$, $R_2=[1,2)$, $R_3=[2,3)$, and $R_4=[3,\infty)$, where the time-dependent covariates are constant within each interval but can vary between intervals. We generated $\bZ=(\bZ\subind\trans,\bZ{\subdep}(t\in R_1)\trans,\bZ{\subdep}(t\in R_2)\trans,\bZ{\subdep}(t\in R_3)\trans,\bZ{\subdep}(t\in R_4)\trans)\trans$ from a multivariate normal distribution with mean $\mathbf{0}_{p\subind+4p\subdep}\trans$ and variance-covariance matrix $\Vbb=[I(l=l')+vI(l \ne l')]_{l=1,...,p\subind+4p\subdep}^{l'=1,...,p\subind+4p\subdep}$, where $\bZ\subind$ are the time-independent covariates and $\bZ{\subdep}(t\in R_\jmath)$ are the time-dependent for $t\in R_\jmath$. We similarly considered $v=0.2, 0.5, 0.8$ to represent weak, moderate, and strong correlations.

We generated $T_i$ from a Weibull distribution with a shape parameter of 2 and a scale parameter of $\{0.05\exp(\bbeta_0\trans \bZ_i(t))\}^{-0.5}$, where 
$\bbeta_0^{\sf(IV)} = (\bbeta\subindt^{\sf(IV)\trans},\bbeta\subdept^{\sf(IV)\trans})\trans$, 
\begin{align*}
\bbeta\subindt^{\sf(IV)}=(\mathbf{0.08}_{3}\trans, \mathbf{0.04}_{3}\trans,\mathbf{0.02}_{3}\trans,\mathbf{0}_{p{\subind}-9}\trans)\trans,\text{ and}\quad
\bbeta\subdept^{\sf(IV)}=(\mathbf{0.08}_{3}\trans, \mathbf{0.04}_{3}\trans,\mathbf{0.02}_{3}\trans,\mathbf{0}_{p{\subdep}-9}\trans)\trans.
\end{align*}
We considered an administrative censoring with $C_i=4$, leading to $44\%$ censoring under the three scenarios represented by weak, moderate, and strong correlations of the design matrix.

\subsubsection{Measures of performance} For any $\bbetahat\in\{\bbetahat\subDAC, \bbetahat\subfull,\bbetahat\subfulllin, \bbetahat\subMV\}$, we report (a) the average computation time for $\bbetahat$; (b) the global mean squared error (GMSE), defined as $(\bbetahat - \bbeta_0)\trans\Vbb(\bbetahat-\bbeta_0)$; (c) empirical probability of $\jmath\not\in\Aschat$; (d) the bias of each individual coefficient; and (e) mean squared error (MSE) of each individual coefficient. For $\bbetahat\subDAC$ and $\bbetahat\subfulllin$, we also report the empirical coverage level of the 95\% normal confidence interval with standard error estimated as described in section \ref{sec-tuning}. For any $\bbetatilde\in\{\bbetatilde\subDAC,\bbetatilde_{\Omega\subfull}\}$, we report (a) the average computation time for $\bbetatilde$; (b) the global mean squared error (GMSE), defined as $(\bbetatilde - \bbeta_0)\trans\Vbb(\bbetatilde-\bbeta_0)$. 

The average computation time for each configuration is based on simulations using 50 simulated datasets performed on Intel\textsuperscript{\textregistered} Xeon\textsuperscript{\textregistered} E5-2620 v3 @2.40GHz. The statistical performance is evaluated based on 1000 simulated datasets for each configuration. 

\subsection{Simulation Results}
We first show in Table \ref{TDCCox} for the time-independent settings and in Table \ref{TDCTVCCox} for the time-dependent settings the average computation time and GMSE of unpenalized estimators $\bbetatilde\subDAC$ and $\bbetatilde_{\Omega\subfull}$. The results suggest that $\bbetatilde\subDAC$ with two iterations ($\Isc=2$) attains a GMSE comparable to the full sample-based estimator $\bbetatilde_{\Omega\subfull}$ and reduced the computation time by more than 50\%. The DAC estimator $\bbetatilde\subDAC$ with two iterations ($\Isc=2$) has a similar GMSE to $\Isc=3$. Across all settings, the results of $\bbetahat\subDAC$ are nearly identical with $\Isc=2$ or $\Isc=3$ and hence we summarize below the results for $\bbetahat\subDAC$ only for $\Isc=2$ unless noted otherwise.
 
\subsubsection{Computation Time}
There are substantial differences in computation time across methods (Table \ref{Tbeta1}-\ref{Tbeta3p200}) for time-independent survival data. Across different settings, the average computation time of $\bbetahat\subDAC$ ranges from $9.6$ to $16.6$ seconds for $p=50$ and from $135.7$ to $181.6$ seconds for $p=200$, with virtually all time spent on the computation of the unpenalized estimator $\bbetatilde\subDAC$. On the contrary, $\bbetahat\subfull$ requires a substantially longer computation time with average time ranging from $409.6$ to $515.3$ seconds for $p=50$ and from $1435.0$ to $1684.9$ seconds for $p=200$. This suggests that the computation time of $\bbetahat\subDAC$ is about $2-3\%$ of the full sample estimator when $p=50$ and about $10\%$ when $p=200$. On the other hand, $\bbetahat\subMV$ has a substantially longer average computation time than $\bbetahat\subfull$. This is because the MCP procedure, requiring more computational time than the adaptive LASSO, needs to be fitted $K=100$ times.

In the presence of time-dependent covariates, Table \ref{Tbeta4} shows that $\bbetahat\subDAC$ has an average computation time of $112.3-120.8$ seconds for $p\subind=50$ and $p\subdep=50$; $\bbetahat\subfulllin$ has an average computation time of $253.9-263.5$ seconds. Virtually all computation time for $\bbetahat\subDAC$ and $\bbetahat\subfulllin$ is spent on the computation of the unpenalized initial estimator $\bbetatilde\subDAC$, which has more observations and requires substantially more computation time compared to the setting with time-independent covariates given the same $n_0$ and $p$.

\subsubsection{Statistical Performance}
The results for the simulation scenarios with only time-independent covariates are summarized in Table \ref{Tbeta1}-\ref{Tbeta3p200}. In general, $\bbetahat\subDAC$ is able to achieve a statistical performance comparable to $\bbetahat\subfull$, while $\bbetahat\subMV$ generally has a worse performance, with respect to the GMSE and variable selection, bias, and MSE of individual coefficient.  For example, as shown in Table \ref{Tbeta1},  the GMSEs ($\times 10^{-5}$) for $\bbetahat\subDAC$, $\bbetahat\subfull$ and $\bbetahat\subMV$ are respectively $4.27$, $4.24$, and $5.61$ when $p=50$ and $v=0.2$; $4.1$, $4.08$, $5.5$ when $p=200$ and $v=0.2$. The relative performance of different procedures has similar patterns across different levels of correlation $v$ among the covariates. When the signals are relatively strong and sparse  as for $\bbeta_0 = \bbeta_0^{\sf{(I)}}$ or $\bbeta_0^{\sf{(II)}}$, $\bbetahat\subDAC$ and $\bbetahat\subfull$ have small biases and achieved perfect variable selection, while $\bbetahat\subMV$ substantially excludes the $\beta_{0j} = 0.05$ signal when $p=200$. For the more challenging case of $\bbeta_0 = \bbeta_0^{\sf{(III)}}$ where some of the signals are weak, the variable selection of $\bbetahat\subDAC$ and $\bbetahat\subfull$ is also near perfect. Both penalized estimators for the weakest signal (0.035) exhibits a small amount of bias when $v = 0.2$ and $0.5$ and an increased bias when $v=0.8$. Such biases in the weak signals are expected for shrinkage estimators \citep{Menelaos2016}, especially in the presence of high correlation among covariates. However, it is important to note that $\bbetahat\subDAC$ and $\bbetahat\subfull$ perform nearly identically, suggesting that our $\DAC\sublin$ procedure incur negligible additional approximation errors. On the other hand, $\bbetahat\subMV$ has difficulty in detecting the 0.05 and 0.035 signals and tends to produce substantially higher MSE than $\bbetahat\subDAC$. 
The empirical coverage levels for the confidence intervals are close to the nominal level across all settings except for the very challenging set very weak signals when the correlation is $v = 0.8$. This again is due to the bias inherent in shrinkage estimators..

The results for the time-dependent survival are summarized in Table \ref{Tbeta4}. We find that $\bbetahat\subDAC$ also generally has a good performance in estimating $\bbeta_0^{\sf{(IV)}}$ for both time-independent and time-dependent covariates. The variable selection consistency holds perfectly for all parameters of interest. The coverage of the confidence intervals also has similar patterns as the case with time-independent covariates.

\section{Application of the DAC procedure to Medicare Data}
\label{sec-app}

We applied the proposed $\DAC\sublin$ algorithm to develop risk prediction models for heart failure-specific readmission or death within 30 days of discharge among Medicare patients who were admitted due to heart failure. The Medicare inpatient claims were assembled for all Medicare fee-for-service beneficiaries during $2000-2012$ to identify the eligible study population. The index date was defined as the discharge date of the first heart failure admission of each patient. We restricted the study population to patients who were discharged alive from the first heart failure admission. The outcome of interest is time to heart failure-specific readmission or death after the first heart failure admission. Because readmission rates within 30 days have been used to assess the quality of care at hospitals by the Centers for Medicare and Medicaid Services (CMS) (\citealp{CMS2017}), we censored the time to readmission at 30 days. For a patient who were readmitted or dead on the same day as discharge (whose claim did not indicate discharge dead), the time-to-event was set at 0.5 days. Due to the large number of ICD-9 codes, we classified each discharge ICD-9 code into disease phenotypes indexed by phenotype codes according to \cite{Denny2013}. A heart failure admission or readmission was identified, if the claim for that admission or readmission had a heart failure phenotype code at discharge. 

We consider two sets of covariates: (I) time-independent covariates including baseline individual-level covariates collected at time of discharge from the index heart failure hospitalization, baseline area-level covariates at the residential ZIP code of each patient, and indicators for time trend including include dummy variables for each year and dummy variables for each months, and (II) time-dependent predictors that vary day-by-day. Baseline individual-level covariates include age, sex, race (white, black, others), calendar year and month of the discharge, Charlson co-morbidity index (CCI) \citep{Quan2005} which describes the degree of illness of a patient, and indicators for non-rare co-morbidities (defined as prevalence $>0.1$ among the study population). Baseline area-level covariates include socioeconomic status variables [percent black residents (ranging from 0 to 1), percent Hispanic residents (ranging from 0 to 1), median household income (per ten thousand increase), median home value (per ten thousand increase), percent below poverty (ranging from 0 to 1), percent below high school (ranging from 0 to 1), percent owned houses (ranging from 0 to 1)], population density (1000 per squared kilometer), and health status variables [percent taking hemoglobin A1C test (ranging 0-1), average BMI, percent ambulance use (ranging from 0 to 1), percent having low-density lipoprotein test (ranging from 0 to 1), and smoke rate (ranging from 0 to 1)]. The time-dependent covariates include daily fine particulate matter (PM$_{2.5}$) predicted using a neural network algorithm  \citep{Di2016}, daily temperature with its quadratic form, and daily dew point temperature with its quadratic form. There were 574 time-independent covariates and 5 time-dependent covariates.

There were $n_0=9,567,752$ eligible patients with a total of $d_0=2,079,436$ heart failure readmissions or deaths, among which $1,453,627$ were readmissions and $625,809$ were deaths. After expanding the dataset by accounting for time-dependent variables which vary day-by-day, the time-dependent dataset contains $245,623,834$ rows of records.

We fit cause-specific Cox models for readmission due to heart failure or deaths as a composite outcome, considering two separate models: (i) a model containing only time-independent covariates and (ii) a model incorporating time-dependent covariates. In both cases, the datasets are too large for \texttt{glmnet} package to analyze as a whole, demonstrating the need for $\DAC\sublin$.

\subsection{Time-independent Covariates Only}
\begin{figure}[!p]
	\centering\includegraphics[height = 580pt]{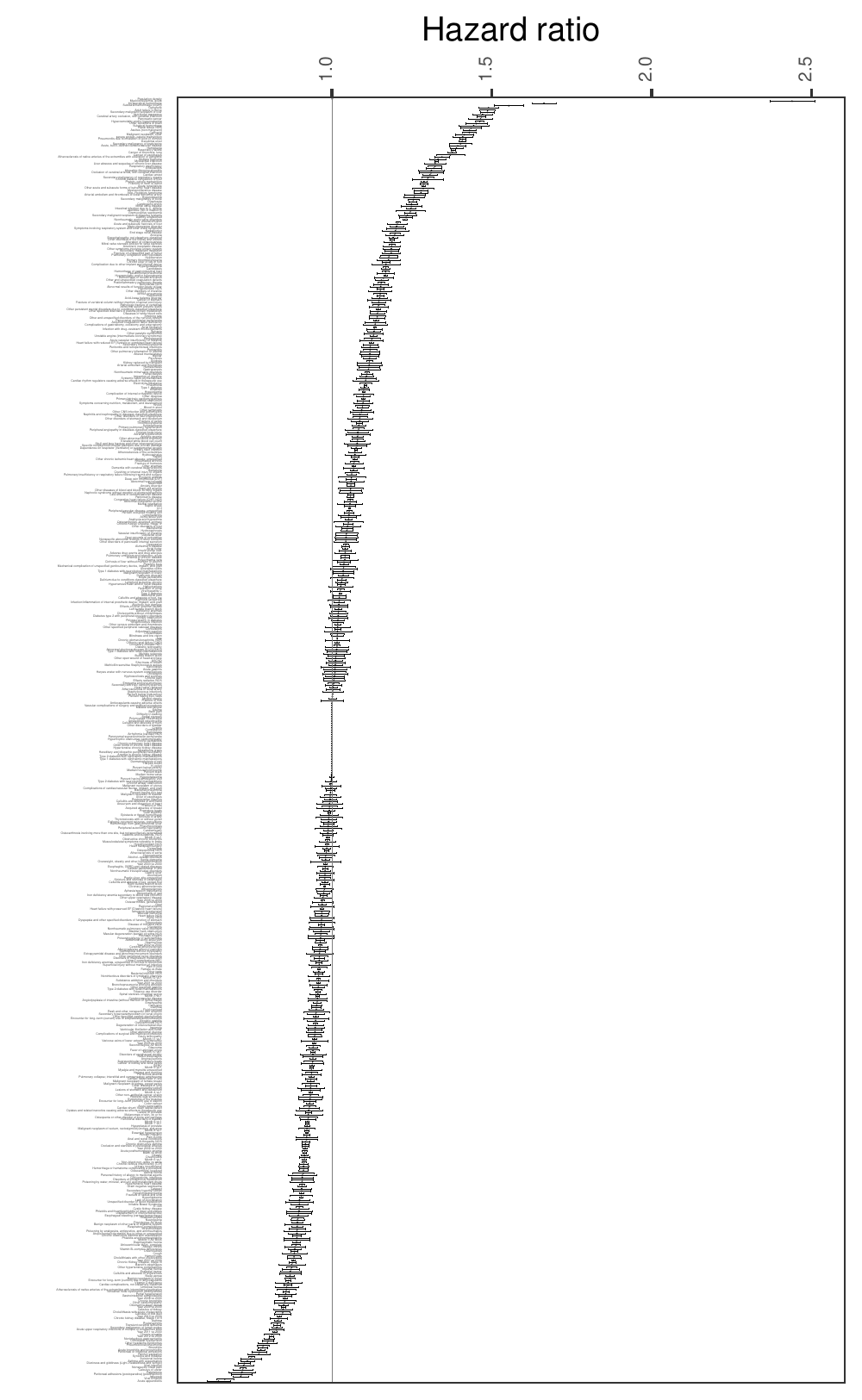}
	\caption{Hazard ratios of each covariate predicting heart failure readmissions or death within 30 days after the first admission using $\DAC\sublin$.}
	\label{Fig1}
\end{figure}

We applied $\DAC\sublin$ with $K=50$ and paralleled $\DAC\sublin$ on 25 Authentic AMD Little Endian @2.30GHz CPUs. Computing $\bbetahat\subDAC$ with $\Isc = 2$ took 1.1 hours, including the time of reading datasets from hard drives during each iteration of the update of the one-step estimator. Figure \ref{Fig1} shows the hazard ratio of each covariate based on $\bbetahat\subDAC$ with $\Isc = 2$ predicting heart failure-specific readmission and death within 30 days. 

Multiple co-morbidities were associated with an increased risk of 30-day readmission or death with the leading factors including renal failures, cancers, malnutrition, subdural or intracerebral hemorrhage, myocardial infarction, endocarditis, respiratory failure, and cardiac arrest. CCI was also associated with an increased hazard of the outcome. These findings are generally consistent with those reported in the literature. For example, \cite{Philbin1999} reported that ischemic heart disease, diabetes, renal diseases, and idiopathic cardiomyopathy were associated with an increased risk of heart failure-specific readmission within a year. Leading factors negatively associated with readmissions included virus infections, asthma, and chronic kidney disease in earlier stages. These negative association findings are reflective of both clinical practice patterns and the biological effects, as most of the negative predictors are generally less severe than the positive predictors. 

Some socioeconomic status predictors were relatively less important in predicting the outcome after accounting for the phenotypes, where percent black, median household income, and percent below poverty were dropped and dual eligibility, median home value, percent below high school had a small hazard ratio. By comparison, \cite{Philbin2001} reported a decrease in readmission as neighborhood income increased. \cite{Foraker2011} reported that given co-morbidity measured by CCI, the readmission or death hazard was higher for low socioeconomic status patients. The present paper considered more detailed phenotypes in addition to CCI suggested a relatively smaller impact of socioeconomic status. The difference in results is possibly because co-morbidity may be on the causal pathway between socioeconomic status and readmission or death. Adjusting for a detailed set of co-morbidities partially blocks the effect of socioeconomic status. Percent Hispanic residents was negatively associated with readmission or death. Percent occupied houses increased the risk of readmission or death, which was consistent with the strong positive prediction by population density. Most ecological health variable showed a small hazard ratio.

Black and other race groups had a lower hazard than white. Females had a lower hazard than males, which was consistent with \cite{Roger2004} that females had a higher survival rate than males after heart failure. Age was associated with an increased hazard of readmission or death, as expected.

The coefficient by month suggested a higher risk of readmission or death in cold seasons than warm seasons, with a larger negative hazard ratio for summer indicators. The short-term readmission or death rate was decreasing over time, which was suggested by the negative hazard ratio of later years. The later calendar year being negatively associated with readmission risk may be an indication of improved follow-up care for patients discharged from heart failure. Consistently, \citep{Roger2004} also suggests an improved heart failure survival rate over time.

\subsection{Incorporating Time-Dependent Covariates}
The analysis has two goals. First, the covariates serve as the risk predictors of the hazard of heart failure-specific readmission. Second, all covariates other than PM$_{2.5}$ serve as the potential confounders of the association between PM$_{2.5}$ and readmission, particularly time trend and area-level covariates. The $\DAC\sublin$ procedure is a variable selection technique to drop non-informative confounders given the high dimensionality of confounders. This goal aligns with \cite{Belloni2014}, which constructs separate penalized regressions for the propensity score model and outcome regression model to identify confounders. We herein focused on building a penalized regression for the outcome regression model.

\begin{figure}[!p]
	\centering\includegraphics[height = 580pt]{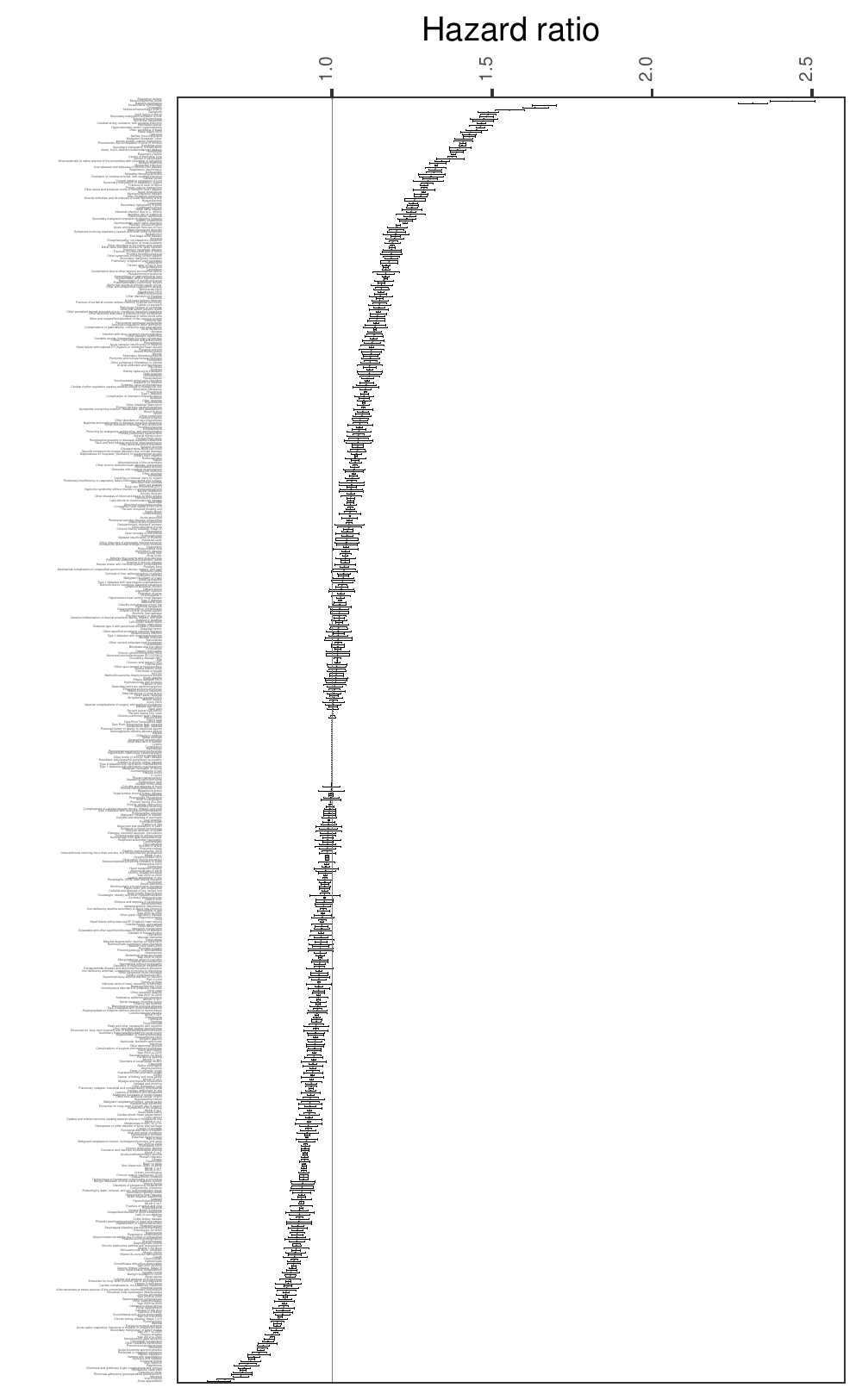}
	\caption{Hazard ratios of each covariate in estimating hazard of heart failure readmissions or death within 30 days associated with PM$_{2.5}$ after the first admission using $\DAC\sublin$.}
	\label{Fig2}
\end{figure}

We applied $\DAC\sublin$ algorithm with $K=200$ to this time-dependent survival dataset. The procedure was paralleled on 10 Authentic AMD Little Endian @2.30GHz CPUs. 
The estimation of $\bbetahat\subDAC$ with $\Isc=2$ took 36.5 hours, including the time of loading the datasets into memory. The result suggests each 10 $\mu g\ m^{-3}$ increase in daily PM$_{2.5}$ was associated with  $0.5\%$ increase of risk (95\% confidence interval $[0.3\%,\ 0.7\%]$) adjusting for individual-level, area-level covariates, and temperature. Because there is rare evidence on whether air pollution is associated with heart failure-specific readmission or death among heart failure patients and it is rare to estimate the health effect of daily air pollution using a time-dependent Cox model, this model provides a novel approach to address a new research question. While evidence is rare on the association between daily PM$_{2.5}$ and heart failure-specific readmission, some studies used case-crossover design to estimate the effect of short-term PM$_{2.5}$ on the incidence of heart failure admissions. \cite{Pope2008} found that a 10 $\mu g\ m^{-3}$ increase in 14-day moving average PM$_{2.5}$ was associated with a 13.1\% ($1.3\%,\ 26.2\%$) increase in the incidence of heart failure admissions among elderly patients; \cite{Zanobetti2009} reported that each 10 $\mu g\ m^{-3}$ increase in 2-day averaged PM$_{2.5}$ was associated with a 1.85\% ($1.19\%,\ 2.51\%$) increase in the incidence of congestive heart failure admission. There is also a large body of literature suggesting that short-term exposure to PM$_{2.5}$ is associated with an increased risk of death. For example, \citep{Di2017} shows among the Medicare population during $2000-2012$ that each 10 $\mu g\ m^{-3}$ increase in PM$_{2.5}$ was associated with an 1.05\% (0.95\%, 1.15\%) increase in mortality risk. In addition, Figure \ref{Fig2} shows the covariate-specific estimates of the hazard ratio for all the covariates, with the estimates consistent with the analysis of time-independent dataset.

\section{Discussions}
\label{sec-discuss}
The proposed $\DAC\sublin$ procedure for fitting adaptive LASSO penalized Cox model reduces the computation cost, while it maintains the precision of estimation and accuracy in variable selection with an extraordinarily large $n_0$ and a numerically large $p$.
The use of $\bbetatilde\subDAC$ makes it feasible to obtain the $\sqrt{n_0}$-consistent estimator required by penalized step (e.g. when there is a constraint in RAM) and shortens the computation time of the initial estimator by $>50\%$. The improvement in the computation time was substantial in the regularized regression step. The LSA converted the fitting of regularized regression from using a dataset of size $n_0$ to a dataset of size $p$. %The precision of $\bbetahat\subDAC$ primarily depends on how well \eqref{PEN} approximates \eqref{penLP}, which is determined by the precision of $\bbetatilde\subDAC$. Given an extremely large sample size, the precision of $\bbetatilde\subDAC$ is guaranteed. For example, in the simulation setting, $\bbetatilde\subDAC$ performs similarly to $\bbetatilde\subfull$.

The majority voting-based method $\bbetahat\subMV$ with MCP \citep{Chen2014} had a substantially longer computation time than $\bbetahat\subfull$. The difference primarily comes from (a) the fact that the Cox model with MCP is fitted $K$ times and (b) the computation efficiency between \texttt{glmnet} algorithm which is more efficient than the MCP algorithm in \texttt{ncvsurv} \citep{Breheny2011}.
 
The difference in variable selection between $\bbetahat\subDAC$ and $\bbetahat\subMV$ \citep{Chen2014} is primarily due to the majority voting. The simulations in \cite{Chen2014} have shown that an increase in the percentage for the majority voting decreased the sensitivity and increased the specificity of variable selection. Similarly in the simulations of the present study with a $50\%$ of the majority vote, Chen and Xie's procedure showed a high specificity but the sensitivity was low for weaker signals as demonstrated in the simulation studies. 

For non-weak signals, the oracle property appears to hold well as evidenced by the simulation results for $\bbeta_0^{\sf(I)}$ and $\bbeta_0^{\sf(II)}$ shown in Tables \ref{Tbeta1}-\ref{Tbeta2}. For weak signals such as 0.035 in $\bbeta_0^{\sf(III)}$, the oracle property does not appear to hold even with $n_0 = 1,000,000$ and the bias in the shrinkage estimators results in confidence intervals with low coverage. This is consistent with the  the impossibility result shown in \cite{Poetscher2009}, which suggests difficulty in deriving precise interval estimators when adaptive LASSO penalty is employed. 

\section*{Acknowledgments}
The study was partially supported by an USEPA grant
RD-83587201, an NIH grant U54 HG007963, and NIEHS grants R01 ES024332 and P30 ES000002. Its contents are solely the responsibility of the grantee and do not necessarily represent the official views of the USEPA. Further, USEPA does not endorse the purchase of any commercial products or services mentioned in the publication. The analysis of the Medicare data was performed on the secured clusters of Research Computing, Faculty of Arts and Sciences at Harvard University.\\
{\it Conflict of Interest}: None declared.

\section{Appendix}
Throughout we assume all regularity conditions required in \cite{Zhang2007}, $p$ is fixed and $\bbeta_0$ belongs to a compact support. 
We next establish the asymptotic equivalence of $\bbetatilde\subDAC$ and $\bbetatilde_{\Omega\subfull}$ in that $\nhalf_0(\bbetatilde\subDAC - \bbetatilde_{\Omega\subfull}) = o_p(1)$. To this end, we note that $\|\bbetatilde_{\Omega_1} - \bbeta_0\| = O_p(\nnhalf)$, 
$\sup_{t, \bbeta}|\widehat\Ssc_{r,\Omega_k}(t,\bbeta) - \Ssc_{r}(t,\bbeta)| = O_p(\nnhalf)$, and
$\sup_{t, \bbeta}|\widehat\Ssc_{r,\sDAC}(t,\bbeta)- \Ssc_{r}(t,\bbeta)| = O_p(\nnhalf_0)$, where $\widehat\Ssc_{r,\sDAC}(t,\bbeta) = 
\Kinv\sumkK \widehat\Ssc_{r,\Omega_k}(t,\bbeta)$. It follows that  $\sup_{\bbeta}\|\Abbhat\subDAC - \Abb(\bbeta)\| = O_p(\nnhalf_0)$. 
From a taylor series expansion, it's straightforward to see that for $k = 2, ..., K$,
$$
\bbetatilde_{\Omega\subklin} - \bbetatilde_{\Omega_k} = O_p(\ninv) = o_p(\nnhalf_0).
$$
On the other hand, $\bbetatilde_{\Omega_k} - \bbeta_0 = \Abb(\bbeta_0)^{-1}\bUhat_{\Omega_k}(\bbeta_0) + O_p(\ninv)$. Therefore, 
$$
\bbetatilde_{\Omega\subklin} - \bbeta_0 = \Abb(\bbeta_0)^{-1}\bUhat_{\Omega_k}(\bbeta_0) + O_p(\ninv).
$$
Furthermore, from the convergence rate of $\widehat\Ssc_{r,\Omega_k}(t,\bbeta)$ and the fact that $\sup_t|\ninv\sum_{i \in \Omega_k} M_i(t, \bbeta_0)|=O_p(\nnhalf)$, we have 
$\bUhat_{\Omega_k}(\bbeta_0) - \bUtilde_{\Omega_k}(\bbeta_0) = O_p(\ninv)$, where
$$
\bUtilde_{\Omega}(\bbeta_0) = \ninv_{\Omega} \sum_{i \in \Omega} \int \left\{\bZ_i(t) - \frac{\Ssc_1(t, \bbeta_0)}{\Ssc_0(t,\bbeta_0)} \right\} d M_i(t, \bbeta_0).
$$
It follows that 
$$
\bbetatilde_{\Omega\subklin} - \bbeta_0 = \Abb(\bbeta_0)^{-1}\bUtilde_{\Omega_k}(\bbeta_0) + O_p(\ninv).
$$
and therefore
\begin{align*}
\bbetatilde\subDAC^{[1]}  - \bbeta_0 & = \Abb(\bbeta_0)^{-1}\Kinv\sumkK \bUtilde_{\Omega_k}(\bbeta_0)  + O_p(\ninv) 
 = \Abb(\bbeta_0)^{-1} \bUtilde_{\Omega\subfull}(\bbeta_0)  + o_p(\nnhalf) . 
\end{align*}
Similarly, we may show that
\begin{align*}
\bbetatilde_{\Omega\subfull}  - \bbeta_0 & = \Abb(\bbeta_0)^{-1}\bUhat_{\Omega\subfull}(\bbeta_0)  + O_p(\ninv_0) 
 = \Abb(\bbeta_0)^{-1} \bUtilde_{\Omega\subfull}(\bbeta_0)  + o_p(\nnhalf_0) . 
\end{align*}
This implies that $\nhalf_0(\bbetatilde\subDAC^{[1]} - \bbetatilde_{\Omega\subfull}) = o_p(1)$. Similar arguments can be
used to show that the equivalence holds for further iterations of $\bbetatilde\subDAC^{[\Isc]}$ with $\Isc \ge 2$. Although
the asymptotic equivalence holds even for $\Isc = 1$, we find that $\Isc=2$ tends to give better approximation in finite
samples.

\bibliographystyle{biorefs}
\bibliography{refs}

% Table 1 - Compare divide-and-conquer one-step estimator with Standard Cox full-sample

\begin{table}[!p]
	\tblcaption{Comparisons of $\bbetatilde\subDAC\ \Isc=1,2,3$ and $\bbetatilde_{\Omega\subfull}$ with respect to average computation time in seconds and global mean squared error (GMSE $\times 10^{-5}$) for the estimation of $\bbeta_0^{\sf(I)}$, $\bbeta_0^{\sf(II)}$, and $\bbeta_0^{\sf(III)}$ using time-independent survival data.

	\label{TDCCox}}

	\centering
	\begin{tabular}{@{}clccccccc@{}}
		\hline
		&                 && \multicolumn{2}{c}{$\bbeta_0=\bbeta_0^{\sf(I)}$} & \multicolumn{2}{c}{$\bbeta_0=\bbeta_0^{\sf(II)}$} 
		& \multicolumn{2}{c}{$\bbeta_0=\bbeta_0^{\sf(III)}$}  \\
		$v$ & Estimator & & Time & GMSE & Time & GMSE & Time & GMSE\\
		\hline
		\hline\hline
		$p = 50$ &&&&&&&&\\
		\hline
		0.2 & & $\Isc=1$ &  5.9 & 19.32 &  5.0 & 21.10 &  5.4 & 21.20 \\ 
		& $\bbetatilde\subDAC$ & $\Isc=2$  & 11.5 & 19.29 &  9.6 & 21.10 & 10.5 & 21.20 \\ 
		&  & $\Isc=3$   & 17.1 & 19.29 & 14.2 & 21.10 & 15.6 & 21.20 \\ 
		& $\bbetatilde_{\Omega\subfull}$ & & 26.5 & 19.21 & 22.6 & 21.03 & 24.9 & 21.13 \\ 
		\hline
		0.5 & & $\Isc=1$ &  5.6 & 18.10 &  6.3 & 19.51 &  5.7 & 20.40 \\ 
		& $\bbetatilde\subDAC$ & $\Isc=2$  & 10.8 & 18.06 & 12.2 & 19.49 & 11.0 & 20.39 \\ 
		&  & $\Isc=3$ & 16.0 & 18.06 & 18.2 & 19.49 & 16.4 & 20.39 \\ 
		& $\bbetatilde_{\Omega\subfull}$ & & 29.1 & 17.99 & 31.5 & 19.44 & 25.6 & 20.35 \\ 
		\hline
		0.8 & & $\Isc=1$ &  6.6 & 17.91 &  6.6 & 18.69 &  5.7 & 19.73 \\ 
		& $\bbetatilde\subDAC$ & $\Isc=2$  & 12.8 & 17.83 & 12.9 & 18.66 & 11.2 & 19.72 \\ 
		&  & $\Isc=3$ & 19.0 & 17.83 & 19.1 & 18.66 & 16.6 & 19.72 \\ 
		& $\bbetatilde_{\Omega\subfull}$ & & 33.3 & 17.74 & 32.7 & 18.60 & 29.2 & 19.65 \\ 
		\hline
		\hline
		$p = 200$&&&&&&&&\\
		\hline
		0.2 & & $\Isc=1$  &  79.1 & 74.94 &  73.2 & 83.81 &  69.4 & 85.02 \\ 
		& $\bbetatilde\subDAC$ & $\Isc=2$  & 155.1 & 74.48 & 143.3 & 83.36 & 135.5 & 84.61 \\ 
		&  & $\Isc=3$ & 231.2 & 74.48 & 213.9 & 83.36 & 201.4 & 84.61 \\ 
		& $\bbetatilde_{\Omega\subfull}$ & & 377.7 & 74.29 & 300.9 & 83.16 & 284.9 & 84.41 \\ 
		\hline
		0.5 & & $\Isc=1$ &  88.8 & 68.99 &  81.3 & 76.47 &  72.1 & 80.63 \\ 
		& $\bbetatilde\subDAC$ & $\Isc=2$  & 173.8 & 68.31 & 158.8 & 76.06 & 141.4 & 80.25 \\ 
		&  & $\Isc=3$ & 258.9 & 68.31 & 236.3 & 76.06 & 210.7 & 80.25 \\ 
		& $\bbetatilde_{\Omega\subfull}$ & & 415.8 & 68.12 & 387.6 & 75.86 & 299.8 & 80.07 \\ 
		\hline
		0.8 & & $\Isc=1$ &  92.9 & 65.96 &  86.3 & 72.03 &  76.0 & 77.25 \\ 
		& $\bbetatilde\subDAC$ & $\Isc=2$  & 181.5 & 65.02 & 168.7 & 71.53 & 148.6 & 76.85 \\ 
		&  & $\Isc=3$ & 269.3 & 65.03 & 251.3 & 71.53 & 221.1 & 76.85 \\ 
		& $\bbetatilde_{\Omega\subfull}$ & & 485.9 & 64.85 & 405.1 & 71.34 & 357.4 & 76.65 \\ 
		\hline\hline

	\end{tabular}
\end{table}

\begin{table}[!p]
	\tblcaption{Comparisons of $\bbetatilde\subDAC\ \Isc=1,2,3$ and $\bbetatilde_{\Omega\subfull}$ with respect to average computation time in seconds and global mean squared error (GMSE $\times 10^{-5}$) for the estimation of $\bbeta_0^{\sf(IV)}$ using time-dependent survival data.
		
		\label{TDCTVCCox}}
	
	\centering
	\begin{tabular}{@{}clccc@{}}
		\hline
		&                 && \multicolumn{2}{c}{$\bbeta_0=\bbeta_0^{\sf(IV)}$}   \\
		$v$ & Estimator & & Time & GMSE \\
		\hline
		\hline\hline
		0.2 & & $\Isc=1$ &  62.0 & 17.94 \\ 
		& $\bbetatilde\subDAC$ & $\Isc=2$  &  120.8 & 17.93 \\ 
		&  & $\Isc=3$  &  178.7 & 17.93 \\ 
		& $\bbetatilde_{\Omega\subfull}$ & & 262.1 & 17.95 \\ 
		\hline 
		0.5 & & $\Isc=1$ &  58.0 & 17.88 \\ 
		& $\bbetatilde\subDAC$ & $\Isc=2$  & 112.2 & 17.88 \\ 
		&  & $\Isc=3$  & 166.3 & 17.88 \\ 
		& $\bbetatilde_{\Omega\subfull}$ & & 263.4 & 17.88 \\ 
		\hline
		0.8 & & $\Isc=1$ &  58.3 & 17.98 \\ 
		& $\bbetatilde\subDAC$ & $\Isc=2$  & 113.8 & 17.98 \\ 
		&  & $\Isc=3$  & 168.5 & 17.98 \\ 
		& $\bbetatilde_{\Omega\subfull}$ & & 253.8 & 17.94 \\ 
		\hline
	\end{tabular}
\end{table}

\begin{table}[!p]
\tblcaption{Comparisons of $\bbetahat\subDAC\ (\Isc=1,2,3)$, $\bbetahat\subfull$, and $\bbetahat\subMV$ for estimating $\bbeta_0=\bbeta_0^{\sf(I)}$ with respect to average computation time in seconds, GMSE ($\times 10^{-5}$), coefficient-specific empirical probability (\%) of $j\not\in\widehat\Asc$, bias ($\times 10^{-4}$), MSE ($\times 10^{-5}$), and empirical coverage probability (\%) of the confidence intervals.
	\label{Tbeta1}}
{\tabcolsep=1.5pt
		\begin{tabular}{@{}ccccccccccccccccc@{}}
		\hline
		&  & \multicolumn{5}{c}{$v = 0.2$} & \multicolumn{5}{c}{$v = 0.5$} & \multicolumn{5}{c}{$v = 0.8$}  \\
		&  & \multicolumn{3}{c}{$\bbetahat\subDAC$} & \raisebox{0ex}{$\bbetahat\subfull$} & \raisebox{0ex}{$\bbetahat\subMV$} 
		& \multicolumn{3}{c}{$\bbetahat\subDAC$} & \raisebox{0ex}{$\bbetahat\subfull$} & \raisebox{0ex}{$\bbetahat\subMV$} 
		& \multicolumn{3}{c}{$\bbetahat\subDAC$} & \raisebox{0ex}{$\bbetahat\subfull$} & \raisebox{0ex}{$\bbetahat\subMV$}  \\
		& $\Isc=$  & $1$ & $2$ & $3$  & & & $1$ & $2$ & $3$  & & & $1$ & $2$ & $3$  & & \\ 
		\hline
		$p=50$ &&&&&&&&&&&&&&&&\\
		\hline
		& Time  &    5.9 &   11.6 &   17.1 &  469.5 & 2140.6 &    5.6 &   10.8 &   16.1 &  485.5 & 1001.0 &    6.7 &   12.9 &   19.0 &  515.3 & 1077.5 \\ 
		& GMSE & 4.32 & 4.27 & 4.27 & 4.24 & 5.61 & 4.53 & 4.42 & 4.42 & 4.41 & 7.40 & 5.17 & 5.01 & 5.01 & 4.97 & 9.91 \\ 
		\hline
		$0.8$ & \%zero & 0.0 & 0.0 & 0.0 & 0.0 & 0.0 & 0.0 & 0.0 & 0.0 & 0.0 & 0.0 & 0.0 & 0.0 & 0.0 & 0.0 & 0.0 \\ 
		& Bias & -0.74 &  0.49 &  0.49 &  0.63 & 12.87 & -1.64 & -0.24 & -0.24 & -0.19 & 12.30 & -0.27 &  1.16 &  1.16 &  1.16 & 14.71 \\ 
		& MSE & 0.55 & 0.55 & 0.55 & 0.55 & 0.72 & 0.70 & 0.70 & 0.70 & 0.70 & 0.86 & 1.50 & 1.50 & 1.50 & 1.50 & 1.76 \\ 
		& CovP & 94.9 & 95.7 & 95.7 &   - &  - & 95.0 & 95.3 & 95.3 & - & - & 96.2 & 96.4 & 96.4 & - & - \\ 
		\hline
		0.4 & \%zero & 0.0 & 0.0 & 0.0 & 0.0 & 0.0 & 0.0 & 0.0 & 0.0 & 0.0 & 0.0 & 0.0 & 0.0 & 0.0 & 0.0 & 0.0 \\ 
		& Bias & -1.47 & -0.85 & -0.85 & -0.72 &  5.61 & -0.63 &  0.07 &  0.07 &  0.03 &  6.55 & -0.02 &  0.72 &  0.72 &  0.73 &  8.81 \\ 
		& MSE & 0.49 & 0.48 & 0.48 & 0.48 & 0.52 & 0.55 & 0.55 & 0.55 & 0.55 & 0.60 & 1.48 & 1.48 & 1.48 & 1.46 & 1.60 \\ 
		& CovP & 93.8 & 94.3 & 94.3 &   - &   - & 97.3 & 97.1 & 97.1 &   - &   - & 95.7 & 95.7 & 95.7 &   - &   - \\ 
		\hline
		0.2 & \%zero & 0.0 & 0.0 & 0.0 & 0.0 & 0.0 & 0.0 & 0.0 & 0.0 & 0.0 & 0.0 & 0.0 & 0.0 & 0.0 & 0.0 & 0.0 \\ 
		& Bias & -1.39 & -1.09 & -1.09 & -1.18 &  2.75 & -0.82 & -0.49 & -0.49 & -0.44 &  3.88 & -3.37 & -3.02 & -3.02 & -2.87 & -3.18 \\ 
		& MSE & 0.44 & 0.44 & 0.44 & 0.44 & 0.46 & 0.63 & 0.63 & 0.63 & 0.63 & 0.67 & 1.49 & 1.48 & 1.48 & 1.48 & 1.76 \\ 
		& CovP & 95.0 & 94.9 & 94.9 &   - &  - & 95.1 & 95.1 & 95.1 &  - & - & 96.0 & 96.1 & 96.1 & - & - \\ 
		\hline
		0  & \%zero & 100.0 & 100.0 & 100.0 & 100.0 & 100.0 & 100.0 & 100.0 & 100.0 & 100.0 & 100.0 & 100.0 & 100.0 & 100.0 & 100.0 & 100.0 \\ 
		& Bias & 0.00 & 0.00 & 0.00 & 0.00 & 0.00 & 0.00 & 0.00 & 0.00 & 0.00 & 0.00 & 0.00 & 0.00 & 0.00 & 0.00 & 0.00 \\ 
		& MSE & 0.00 & 0.00 & 0.00 & 0.00 & 0.00 & 0.00 & 0.00 & 0.00 & 0.00 & 0.00 & 0.00 & 0.00 & 0.00 & 0.00 & 0.00 \\ 
		\hline\hline
		$p=200$ &&&&&&&&&&&&&&&&\\
		\hline
		& Time &   79.3 &   155.3 &   231.4 &  1539.0 & 21029.7 &   89.0 &  174.0 &  259.0 & 1684.9 & 4256.8 &   93.1 &  181.6 &  269.5 & 1664.5 & 4814.5 \\ 
		& GMSE &4.62 & 4.10 & 4.10 & 4.08 & 5.50 & 5.83 & 4.61 & 4.61 & 4.61 & 9.50 &  6.91 &  5.05 &  5.05 &  5.06 & 17.02 \\ 
		\hline
		0.8 & \%zero & 0.0 & 0.0 & 0.0 & 0.0 & 0.0 & 0.0 & 0.0 & 0.0 & 0.0 & 0.0 & 0.0 & 0.0 & 0.0 & 0.0 & 0.0 \\ 
		& Bias & -5.46 &  0.31 &  0.31 &  0.27 & 13.21 & -6.75 &  0.25 &  0.25 &  0.33 & 16.02 & -6.38 &  0.94 &  0.94 &  0.84 & 20.72 \\ 
		& MSE & 0.58 & 0.53 & 0.53 & 0.53 & 0.71 & 0.77 & 0.70 & 0.70 & 0.70 & 0.97 & 1.49 & 1.45 & 1.45 & 1.45 & 1.96 \\ 
		& CovP & 94.5 & 95.2 & 95.2 & - & - & 95.0 & 96.0 & 96.0 & - & - & 96.3 & 96.2 & 96.2 & - & - \\ 
		\hline
		0.4 & \%zero & 0.0 & 0.0 & 0.0 & 0.0 & 0.0 & 0.0 & 0.0 & 0.0 & 0.0 & 0.0 & 0.0 & 0.0 & 0.0 & 0.0 & 0.0 \\ 
		& Bias & -2.09 &  0.80 &  0.80 &  0.81 &  7.88 & -3.11 &  0.29 &  0.29 &  0.43 &  8.89 & -3.47 &  0.24 &  0.24 &  0.57 & 11.37 \\ 
		& MSE & 0.45 & 0.44 & 0.44 & 0.44 & 0.51 & 0.68 & 0.66 & 0.66 & 0.66 & 0.76 & 1.42 & 1.42 & 1.42 & 1.42 & 1.62 \\ 
		& CovP & 95.5 & 96.2 & 96.2 & - & - & 95.6 & 95.2 & 95.2 & - & - & 96.0 & 95.8 & 95.8 & - & - \\ 
		\hline
		0.2 & \%zero & 0.0 & 0.0 & 0.0 & 0.0 & 0.0 & 0.0 & 0.0 & 0.0 & 0.0 & 0.0 & 0.0 & 0.0 & 0.0 & 0.0 & 0.0 \\ 
		& Bias & -4.15 & -2.68 & -2.68 & -2.69 &  0.40 & -4.15 & -2.44 & -2.44 & -2.52 &  3.63 & -2.02 & -0.15 & -0.15 & -0.09 &  1.09 \\ 
		& MSE & 0.44 & 0.43 & 0.43 & 0.43 & 0.45 & 0.61 & 0.60 & 0.60 & 0.60 & 0.61 & 1.34 & 1.33 & 1.33 & 1.33 & 1.62 \\ 
		& CovP & 96.0 & 96.5 & 96.5 & - & - & 95.5 & 95.8 & 95.8 & - & - & 97.1 & 97.1 & 97.1 & - & - \\ 
		\hline
		0 & \%zero & 100.0 & 100.0 & 100.0 & 100.0 & 100.0 & 100.0 & 100.0 & 100.0 & 100.0 & 100.0 & 100.0 & 100.0 & 100.0 & 100.0 & 100.0 \\ 
		& Bias & 0.00 & 0.00 & 0.00 & 0.00 & 0.00 & 0.00 & 0.00 & 0.00 & 0.00 & 0.00 & 0.00 & 0.00 & 0.00 & 0.00 & 0.00 \\ 
		& MSE & 0.00 & 0.00 & 0.00 & 0.00 & 0.00 & 0.00 & 0.00 & 0.00 & 0.00 & 0.00 & 0.00 & 0.00 & 0.00 & 0.00 & 0.00 \\ 
		\hline
	\end{tabular}
}
\end{table}

\begin{table}[!p]
	\tblcaption{Comparisons of $\bbetahat\subDAC\ (\Isc=1,2,3)$, $\bbetahat\subfull$, and $\bbetahat\subMV$ for estimating $\bbeta_0=\bbeta_0^{\sf(II)}$ with respect to average computation time in seconds, GMSE ($\times 10^{-5}$), coefficient-specific empirical probability (\%) of $j\not\in\widehat\Asc$, bias ($\times 10^{-4}$), MSE ($\times 10^{-5}$), and empirical coverage probability (\%) of the confidence intervals.
		\label{Tbeta2}}
	{\tabcolsep=1.5pt
		\begin{tabular}{@{}ccccccccccccccccc@{}}
			\hline
			&  & \multicolumn{5}{c}{$v = 0.2$} & \multicolumn{5}{c}{$v = 0.5$} & \multicolumn{5}{c}{$v = 0.8$}  \\
			&  & \multicolumn{3}{c}{$\bbetahat\subDAC$} & \raisebox{0ex}{$\bbetahat\subfull$} & \raisebox{0ex}{$\bbetahat\subMV$} 
			& \multicolumn{3}{c}{$\bbetahat\subDAC$} & \raisebox{0ex}{$\bbetahat\subfull$} & \raisebox{0ex}{$\bbetahat\subMV$} 
			& \multicolumn{3}{c}{$\bbetahat\subDAC$} & \raisebox{0ex}{$\bbetahat\subfull$} & \raisebox{0ex}{$\bbetahat\subMV$}  \\
			& $\Isc=$  & $1$ & $2$ & $3$  & & & $1$ & $2$ & $3$  & & & $1$ & $2$ & $3$  & & \\ 
			\hline
			$p=50$ &&&&&&&&&&&&&&&&\\
		\hline
		& Time &    5.0 &    9.6 &   14.2 &  440.1 & 2147.8 &    6.3 &   12.3 &   18.2 &  479.8 & 1223.6 &    6.6 &   12.9 &   19.2 &  505.4 & 1133.0 \\ 
		& GMSE &   7.27 &   7.24 &   7.24 &   7.22 & 869.51 &    6.92 &    6.86 &    6.86 &    6.84 & 1176.14 &    7.21 &    7.12 &    7.12 &    7.17 & 4451.57 \\ 
		\hline
		0.4 & \%zero & 0.0 & 0.0 & 0.0 & 0.0 & 0.0 & 0.0 & 0.0 & 0.0 & 0.0 & 0.0 & 0.0 & 0.0 & 0.0 & 0.0 & 0.0 \\ 
		& Bias &  1.14 &  1.59 &  1.59 &  1.60 & 97.75 &   1.47 &   2.07 &   2.07 &   2.11 & 109.82 &   4.82 &   5.45 &   5.45 &   1.39 & 218.73 \\ 
		& MSE &  0.53 &  0.53 &  0.53 &  0.53 & 10.11 &  0.71 &  0.71 &  0.71 &  0.71 & 12.83 &  1.70 &  1.70 &  1.70 &  1.68 & 49.58 \\ 
		& CovP & 95.1 & 95.1 & 95.1 & - &  - & 95.6 & 95.5 & 95.5 &  - & - & 95.2 & 95.3 & 95.3 & - & - \\ 
		\hline
		0.2 & \%zero & 0.0 & 0.0 & 0.0 & 0.0 & 0.0 & 0.0 & 0.0 & 0.0 & 0.0 & 0.0 & 0.0 & 0.0 & 0.0 & 0.0 & 0.0 \\ 
		& Bias &  0.72 &  0.94 &  0.94 &  1.14 & 98.26 &   1.61 &   1.90 &   1.90 &   2.02 & 107.74 &   3.63 &   3.95 &   3.95 &   1.06 & 205.50 \\ 
		& MSE &  0.50 &  0.50 &  0.50 &  0.50 & 10.19 &  0.76 &  0.76 &  0.76 &  0.75 & 12.43 &  1.65 &  1.65 &  1.65 &  1.66 & 44.21 \\ 
		& CovP & 95.5 & 95.7 & 95.7 &  - &   - & 95.5 & 95.4 & 95.4 &   - &   - & 95.9 & 96.1 & 96.1 &   - &   - \\ 
		\hline
		0.05 & \%zero &  0.0 &  0.0 &  0.0 &  0.0 & 61.6 &   0.0 &   0.0 &   0.0 &   0.0 & 100.0 &   0.0 &   0.0 &   0.0 &   0.0 & 100.0 \\ 
		& Bias &   -4.21 &   -4.16 &   -4.16 &   -4.21 & -406.64 &   -5.65 &   -5.55 &   -5.55 &   -5.52 & -500.00 &  -11.40 &  -11.31 &  -11.31 &   -7.22 & -500.00 \\ 
		& MSE &   0.50 &   0.50 &   0.50 &   0.49 & 179.49 &   0.71 &   0.71 &   0.71 &   0.70 & 250.00 &   2.08 &   2.07 &   2.07 &   2.05 & 250.00 \\ 
		& CovP & 95.2 & 95.2 & 95.2 &  - & -  & 94.8 & 94.9 & 94.9 &   - &   - & 91.5 & 91.7 & 91.7 &   - &   - \\ 
		\hline
		0 & \%zero & 100.0 & 100.0 & 100.0 & 100.0 & 100.0 & 100.0 & 100.0 & 100.0 & 100.0 & 100.0 &  99.9 & 100.0 & 100.0 & 100.0 & 100.0 \\ 
		& Bias & 0.00 & 0.00 & 0.00 & 0.00 & 0.00 & 0.00 & 0.00 & 0.00 & 0.00 & 0.00 & 0.09 & 0.00 & 0.00 & 0.00 & 0.00 \\ 
		& MSE & 0.00 & 0.00 & 0.00 & 0.00 & 0.00 & 0.00 & 0.00 & 0.00 & 0.00 & 0.00 & 0.01 & 0.00 & 0.00 & 0.00 & 0.00 \\ 
		\hline\hline
		$p = 200$ \\ 
		\hline
		& Time &  73.4 &   143.5 &   214.1 &  1435.0 & 20459.8 &   81.5 &  159.0 &  236.5 & 1560.5 & 4646.5 &   86.5 &  168.9 &  251.5 & 1666.2 & 5056.3 \\ 
		& GMSE &   7.53 &    7.29 &    7.29 &    7.28 & 1496.27 &    7.58 &    7.05 &    7.05 &    7.03 & 1300.22 &     8.38 &     7.46 &     7.46 &     7.67 & 12696.30 \\ 
		\hline
		0.4 & \%zero & 0.0 & 0.0 & 0.0 & 0.0 & 0.0 & 0.0 & 0.0 & 0.0 & 0.0 & 0.0 & 0.0 & 0.0 & 0.0 & 0.0 & 0.0 \\ 
		& Bias &  -0.02 &   2.00 &   2.00 &   2.07 & 138.62 &   0.84 &   3.59 &   3.59 &   3.58 & 125.20 &   5.70 &   8.81 &   8.81 &   0.58 & 171.68 \\ 
		& MSE &  0.54 &  0.54 &  0.54 &  0.54 & 19.78 &  0.80 &  0.79 &  0.79 &  0.79 & 16.53 &  1.62 &  1.67 &  1.67 &  1.62 & 31.19 \\ 
		& CovP & 95.4 & 94.9 & 94.9 &   - &  - & 94.8 & 94.9 & 94.9 &  - & - & 96.3 & 96.2 & 96.2 & - & - \\ 
		\hline
		0.2 & \%zero & 0.0 & 0.0 & 0.0 & 0.0 & 0.0 & 0.0 & 0.0 & 0.0 & 0.0 & 0.0 & 0.0 & 0.0 & 0.0 & 0.0 & 0.0 \\ 
		& Bias &  -0.02 &   0.99 &   0.99 &   1.08 & 142.42 &   0.15 &   1.59 &   1.59 &   1.77 & 122.19 &   3.94 &   5.37 &   5.37 &   5.56 & 151.36 \\ 
		& MSE &  0.49 &  0.49 &  0.49 &  0.49 & 20.82 &  0.75 &  0.75 &  0.75 &  0.74 & 15.76 &  1.72 &  1.73 &  1.73 &  1.79 & 25.05 \\ 
		& CovP & 96.4 & 96.4 & 96.4 &   - &   - & 95.4 & 95.2 & 95.2 &   - &   - & 95.1 & 95.3 & 95.3 &   - &   - \\ 
		\hline
		0.05 & \%zero &  0.0 &  0.0 &  0.0 &  0.0 & 99.1 &   0.0 &   0.0 &   0.0 &   0.0 & 100.0 &   0.0 &   0.0 &   0.0 &   0.0 & 100.0 \\ 
		& Bias &   -6.39 &   -6.07 &   -6.07 &   -6.08 & -498.29 &   -8.32 &   -7.92 &   -7.92 &   -7.83 & -500.00 &  -18.45 &  -18.11 &  -18.11 &  -15.39 & -500.00 \\ 
		& MSE &   0.59 &   0.57 &   0.57 &   0.57 & 248.62 &   0.84 &   0.82 &   0.82 &   0.82 & 250.00 &   2.38 &   2.38 &   2.38 &   2.43 & 250.00 \\ 
		& CovP & 93.6 & 93.2 & 93.2 &   - &   - & 93.1 & 93.5 & 93.5 &   - &   - & 90.5 & 91.0 & 91.0 &  - &   - \\ 
		\hline
		0 & \%zero & 100.0 & 100.0 & 100.0 & 100.0 & 100.0 & 100.0 & 100.0 & 100.0 & 100.0 & 100.0 & 100.0 & 100.0 & 100.0 & 100.0 & 100.0 \\ 
		& Bias & 0.00 & 0.00 & 0.00 & 0.00 & 0.00 & 0.00 & 0.00 & 0.00 & 0.00 & 0.00 & 0.00 & 0.00 & 0.00 & 0.00 & 0.00 \\ 
		& MSE & 0.00 & 0.00 & 0.00 & 0.00 & 0.00 & 0.00 & 0.00 & 0.00 & 0.00 & 0.00 & 0.00 & 0.00 & 0.00 & 0.00 & 0.00 \\ 
		\hline
	\end{tabular}}
\end{table}

\begin{table}[!p]
	\tblcaption{Comparisons of $\bbetahat\subDAC\ (\Isc=1,2,3)$, $\bbetahat\subfull$, and $\bbetahat\subMV$ for estimating $\bbeta_0=\bbeta_0^{\sf(III)}$ when $p=50$ with respect to average computation time in seconds, GMSE ($\times 10^{-5}$), coefficient-specific empirical probability (\%) of $j\not\in\widehat\Asc$, bias ($\times 10^{-4}$), MSE ($\times 10^{-5}$), and empirical coverage probability (\%) of the confidence intervals.
		\label{Tbeta3p50}}
	{\tabcolsep=1.5pt
		\begin{tabular}{@{}ccccccccccccccccc@{}}
			\hline
			&  & \multicolumn{5}{c}{$v = 0.2$} & \multicolumn{5}{c}{$v = 0.5$} & \multicolumn{5}{c}{$v = 0.8$}  \\
			&  & \multicolumn{3}{c}{$\bbetahat\subDAC$} & \raisebox{0ex}{$\bbetahat\subfull$} & \raisebox{0ex}{$\bbetahat\subMV$} 
			& \multicolumn{3}{c}{$\bbetahat\subDAC$} & \raisebox{0ex}{$\bbetahat\subfull$} & \raisebox{0ex}{$\bbetahat\subMV$} 
			& \multicolumn{3}{c}{$\bbetahat\subDAC$} & \raisebox{0ex}{$\bbetahat\subfull$} & \raisebox{0ex}{$\bbetahat\subMV$}  \\
			& $\Isc=$  & $1$ & $2$ & $3$  & & & $1$ & $2$ & $3$  & & & $1$ & $2$ & $3$  & & \\ 
			\hline
			$p=50$ &&&&&&&&&&&&&&&&\\
			\hline
		& Time &    5.4 &   10.5 &   15.7 &  416.4 & 1975.2 &    5.7 &   11.1 &   16.4 &  409.6 & 1511.2 &   5.8 &  11.2 &  16.6 & 419.2 & 805.4 \\ 
		& GMSE &  5.36 &    5.35 &    5.35 &    5.33 & 1154.39 &    5.17 &    5.14 &    5.14 &    5.13 & 1261.31 &    5.70 &    5.67 &    5.67 &    5.65 & 3144.68 \\
		\hline
		1 & \%zero & 0.0 & 0.0 & 0.0 & 0.0 & 0.0 & 0.0 & 0.0 & 0.0 & 0.0 & 0.0 & 0.0 & 0.0 & 0.0 & 0.0 & 0.0 \\ 
		& Bias &   1.70 &   2.78 &   2.78 &   2.90 & 149.35 &   2.24 &   3.55 &   3.55 &   3.69 & 177.28 &  11.68 &  13.06 &  13.05 &  11.39 & 261.71 \\ 
		& MSE &  0.67 &  0.68 &  0.68 &  0.68 & 23.00 &  0.91 &  0.91 &  0.91 &  0.91 & 32.44 &  2.07 &  2.10 &  2.10 &  2.05 & 70.53 \\ 
		& CovP & 95.0 & 95.0 & 95.0 &   - &   - & 94.9 & 94.7 & 94.7 &   - &   - & 94.9 & 94.7 & 94.7 &   - &   - \\
		\hline 
		0.5 & \%zero & 0.0 & 0.0 & 0.0 & 0.0 & 0.0 & 0.0 & 0.0 & 0.0 & 0.0 & 0.0 & 0.0 & 0.0 & 0.0 & 0.0 & 0.0 \\ 
		& Bias &   2.54 &   3.06 &   3.06 &   3.14 & 158.85 &   4.90 &   5.53 &   5.53 &   5.71 & 177.87 &  13.53 &  14.22 &  14.22 &  13.60 & 258.65 \\ 
		& MSE &  0.54 &  0.54 &  0.54 &  0.54 & 25.80 &  0.77 &  0.78 &  0.78 &  0.78 & 32.53 &  1.94 &  1.96 &  1.96 &  1.94 & 68.80 \\ 
		& CovP & 95.0 & 95.0 & 95.0 &   - &   - & 94.8 & 94.8 & 94.8 &   - &   - & 94.8 & 94.4 & 94.4 &   - &   - \\ 
		\hline
		0.2 & \%zero & 0.0 & 0.0 & 0.0 & 0.0 & 0.0 & 0.0 & 0.0 & 0.0 & 0.0 & 0.0 & 0.0 & 0.0 & 0.0 & 0.0 & 0.0 \\ 
		& Bias &   2.02 &   2.26 &   2.26 &   2.36 & 162.72 &   2.00 &   2.21 &   2.21 &   2.39 & 174.73 &   7.91 &   8.17 &   8.17 &   8.50 & 244.46 \\ 
		& MSE &  0.55 &  0.55 &  0.55 &  0.54 & 27.10 &  0.73 &  0.73 &  0.73 &  0.73 & 31.42 &  1.91 &  1.91 &  1.91 &  1.92 & 61.97 \\ 
		& CovP & 94.8 & 95.0 & 95.0 &   - &   - & 96.1 & 96.0 & 96.0 &   - &   - & 94.7 & 94.7 & 94.7 &   - &   - \\ 
		\hline
		0.1 & \%zero & 0.0 & 0.0 & 0.0 & 0.0 & 0.0 & 0.0 & 0.0 & 0.0 & 0.0 & 0.0 &  0.0 &  0.0 &  0.0 &  0.0 & 29.1 \\ 
		& Bias &  -0.73 &  -0.60 &  -0.60 &  -0.55 & -66.80 &  1.57 &  1.71 &  1.71 &  1.92 & 58.13 &    5.84 &    5.99 &    5.99 &    6.77 & -390.41 \\ 
		& MSE & 0.51 & 0.51 & 0.51 & 0.51 & 5.75 & 0.65 & 0.65 & 0.65 & 0.65 & 5.25 &   1.84 &   1.85 &   1.85 &   1.86 & 307.51 \\ 
		& CovP & 95.8 & 95.6 & 95.6 &   - &   - & 96.7 & 96.4 & 96.4 &   - &   - & 95.0 & 95.1 & 95.1 &   - &   - \\ 
		\hline
		0.05 & \%zero &  0.0 &  0.0 &  0.0 &  0.0 & 80.1 &   0.0 &   0.0 &   0.0 &   0.0 & 100.0 &   0.0 &   0.0 &   0.0 &   0.0 & 100.0 \\ 
		& Bias &   -3.27 &   -3.22 &   -3.22 &   -3.15 & -451.64 &   -2.92 &   -2.85 &   -2.85 &   -2.77 & -500.00 &   -7.17 &   -7.13 &   -7.13 &   -6.80 & -500.00 \\ 
		& MSE &   0.53 &   0.53 &   0.53 &   0.53 & 213.48 &   0.81 &   0.81 &   0.81 &   0.81 & 250.00 &   2.05 &   2.05 &   2.05 &   2.03 & 250.00 \\ 
		& CovP & 94.2 & 94.1 & 94.1 &   - &   - & 95.0 & 95.0 & 95.0 &   - &   - & 94.0 & 93.9 & 93.9 &   - &   - \\ 
		\hline
		0.035 & \%zero &   0.0 &   0.0 &   0.0 &   0.0 & 100.0 &   0.0 &   0.0 &   0.0 &   0.0 & 100.0 &   0.0 &   0.0 &   0.0 &   0.0 & 100.0 \\ 
		& Bias &   -4.72 &   -4.67 &   -4.67 &   -4.57 & -350.00 &   -6.74 &   -6.66 &   -6.66 &   -6.64 & -350.00 &  -17.13 &  -17.07 &  -17.07 &  -15.98 & -350.00 \\ 
		& MSE &   0.55 &   0.55 &   0.55 &   0.55 & 122.50 &   0.83 &   0.83 &   0.83 &   0.83 & 122.50 &   2.29 &   2.29 &   2.29 &   2.23 & 122.50 \\ 
		& CovP & 94.8 & 94.8 & 94.8 &   - &   - & 94.0 & 93.9 & 93.9 &   - &   - & 92.1 & 92.1 & 92.1 &   - &   - \\ 
		\hline
		0 & \%zero & 100.0 & 100.0 & 100.0 & 100.0 & 100.0 & 100.0 & 100.0 & 100.0 &  99.9 & 100.0 & 100.0 & 100.0 & 100.0 & 100.0 & 100.0 \\ 
		& Bias & 0.00 & 0.00 & 0.00 & 0.00 & 0.00 & 0.00 & 0.00 & 0.00 & 0.06 & 0.00 & 0.00 & 0.00 & 0.00 & 0.00 & 0.00 \\ 
		&MSE & 0.00 & 0.00 & 0.00 & 0.00 & 0.00 & 0.00 & 0.00 & 0.00 & 0.00 & 0.00 & 0.00 & 0.00 & 0.00 & 0.00 & 0.00 \\ 
		\hline
	\end{tabular}}
\end{table}

\begin{table}[!p]
	\tblcaption{Comparisons of $\bbetahat\subDAC\ (\Isc=1,2,3)$, $\bbetahat\subfull$, and $\bbetahat\subMV$ for estimating $\bbeta_0=\bbeta_0^{\sf(III)}$ when $p=200$ with respect to average computation time in seconds, GMSE ($\times 10^{-5}$), coefficient-specific empirical probability (\%) of $j\not\in\widehat\Asc$, bias ($\times 10^{-4}$), MSE ($\times 10^{-5}$), and empirical coverage probability (\%) of the confidence intervals.
		\label{Tbeta3p200}}
	{\tabcolsep=1.5pt
		\begin{tabular}{@{}ccccccccccccccccc@{}}
			\hline
			&  & \multicolumn{5}{c}{$v = 0.2$} & \multicolumn{5}{c}{$v = 0.5$} & \multicolumn{5}{c}{$v = 0.8$}  \\
			&  & \multicolumn{3}{c}{$\bbetahat\subDAC$} & \raisebox{0ex}{$\bbetahat\subfull$} & \raisebox{0ex}{$\bbetahat\subMV$} 
			& \multicolumn{3}{c}{$\bbetahat\subDAC$} & \raisebox{0ex}{$\bbetahat\subfull$} & \raisebox{0ex}{$\bbetahat\subMV$} 
			& \multicolumn{3}{c}{$\bbetahat\subDAC$} & \raisebox{0ex}{$\bbetahat\subfull$} & \raisebox{0ex}{$\bbetahat\subMV$}  \\
			& $\Isc=$  & $1$ & $2$ & $3$  & & & $1$ & $2$ & $3$  & & & $1$ & $2$ & $3$  & & \\ 
			\hline
			$p=200$ &&&&&&&&&&&&&&&&\\
			\hline
			& Time &  69.6 &   135.7 &   201.5 &  1339.0 & 19425.3 &   72.3 &  141.5 &  210.9 & 1381.1 & 3121.6 &   76.2 &  148.7 &  221.3 & 1393.8 & 3788.2 \\ 
			& GMSE &    5.53 &    5.38 &    5.38 &    5.37 & 1438.17 &    5.70 &    5.45 &    5.45 &    5.44 & 1377.06 &    6.68 &    6.27 &    6.27 &    6.25 & 8328.95 \\ 
			\hline
			1 & \%zero & 0.0 & 0.0 & 0.0 & 0.0 & 0.0 & 0.0 & 0.0 & 0.0 & 0.0 & 0.0 & 0.0 & 0.0 & 0.0 & 0.0 & 0.0 \\ 
			& Bias &  -0.78 &   3.84 &   3.84 &   3.90 & 186.92 &   2.03 &   7.72 &   7.72 &   7.90 & 197.72 &  14.18 &  20.69 &  20.69 &  17.61 & 195.98 \\ 
			& MSE &  0.67 &  0.68 &  0.68 &  0.68 & 35.63 &  0.87 &  0.91 &  0.91 &  0.92 & 40.12 &  2.05 &  2.29 &  2.29 &  2.16 & 40.41 \\ 
			& CovP & 95.4 & 95.1 & 95.1 &   - &   - & 95.5 & 94.6 & 94.6 &   - &   - & 95.3 & 94.1 & 94.1 &   - &   - \\ 
			\hline
			0.5 & \%zero & 0.0 & 0.0 & 0.0 & 0.0 & 0.0 & 0.0 & 0.0 & 0.0 & 0.0 & 0.0 & 0.0 & 0.0 & 0.0 & 0.0 & 0.0 \\ 
			& Bias &   2.06 &   4.46 &   4.46 &   4.56 & 205.74 &   5.67 &   8.47 &   8.47 &   8.64 & 197.18 &  17.50 &  20.60 &  20.60 &  19.01 & 186.32 \\ 
			& MSE &  0.50 &  0.51 &  0.51 &  0.51 & 42.85 &  0.77 &  0.81 &  0.81 &  0.81 & 39.75 &  2.19 &  2.28 &  2.28 &  2.20 & 36.67 \\ 
			& CovP & 96.3 & 96.4 & 96.4 &   - &   - & 95.4 & 94.6 & 94.6 &   - &   - & 93.2 & 92.8 & 92.8 &   - &   - \\ 
			\hline
			0.2 & \%zero & 0.0 & 0.0 & 0.0 & 0.0 & 0.0 & 0.0 & 0.0 & 0.0 & 0.0 & 0.0 & 0.0 & 0.0 & 0.0 & 0.0 & 0.0 \\ 
			& Bias &   0.73 &   1.64 &   1.64 &   1.73 & 203.39 &   4.57 &   5.63 &   5.63 &   5.81 & 195.00 &  14.72 &  15.90 &  15.90 &  15.74 & 162.90 \\ 
			& MSE &  0.46 &  0.46 &  0.46 &  0.46 & 41.94 &  0.74 &  0.74 &  0.74 &  0.74 & 38.86 &  1.92 &  1.94 &  1.94 &  1.94 & 28.81 \\ 
			& CovP & 96.8 & 96.6 & 96.6 &   - &   - & 96.1 & 96.1 & 96.1 &   - &   - & 95.3 & 94.5 & 94.5 &   - &   - \\ 
			\hline
			0.1 & \%zero & 0.0 & 0.0 & 0.0 & 0.0 & 0.0 & 0.0 & 0.0 & 0.0 & 0.0 & 0.0 &  0.0 &  0.0 &  0.0 &  0.0 & 88.5 \\ 
			& Bias &    0.77 &    1.30 &    1.30 &    1.43 & -198.61 &   3.40 &   4.01 &   4.01 &   4.34 & -19.37 &    6.65 &    7.27 &    7.28 &    8.22 & -906.50 \\ 
			& MSE &  0.48 &  0.48 &  0.48 &  0.48 & 40.82 & 0.75 & 0.75 & 0.75 & 0.75 & 3.09 &   1.82 &   1.81 &   1.81 &   1.83 & 889.22 \\ 
			& CovP & 96.0 & 95.9 & 95.9 &   - &   - & 95.4 & 95.6 & 95.6 &   - &   - & 94.9 & 95.2 & 95.2 &   - &   - \\ 
			\hline
			0.05 & \%zero &  0.0 &  0.0 &  0.0 &  0.0 & 99.9 &   0.0 &   0.0 &   0.0 &   0.0 & 100.0 &   0.0 &   0.0 &   0.0 &   0.0 & 100.0 \\ 
			& Bias &   -4.72 &   -4.44 &   -4.44 &   -4.35 & -499.81 &   -5.54 &   -5.21 &   -5.21 &   -5.03 & -500.00 &  -11.07 &  -10.82 &  -10.82 &  -10.27 & -500.00 \\ 
			& MSE &   0.51 &   0.50 &   0.50 &   0.50 & 249.85 &   0.81 &   0.80 &   0.80 &   0.80 & 250.00 &   2.17 &   2.15 &   2.15 &   2.14 & 250.00 \\ 
			& CovP & 95.1 & 95.2 & 95.2 &   - &   - & 94.4 & 94.5 & 94.5 &   - &   - & 93.5 & 93.4 & 93.4 &   - &   - \\ 
			\hline
			0.035 & \%zero &   0.0 &   0.0 &   0.0 &   0.0 & 100.0 &   0.0 &   0.0 &   0.0 &   0.0 & 100.0 &   0.0 &   0.0 &   0.0 &   0.0 & 100.0 \\ 
			& Bias &   -9.70 &   -9.55 &   -9.55 &   -9.54 & -350.00 &  -10.90 &  -10.66 &  -10.66 &  -10.63 & -350.00 &  -29.87 &  -29.37 &  -29.37 &  -27.09 & -350.00 \\ 
			& MSE &   0.63 &   0.63 &   0.63 &   0.63 & 122.50 &   0.98 &   0.97 &   0.97 &   0.97 & 122.50 &   3.36 &   3.30 &   3.30 &   3.16 & 122.50 \\ 
			& CovP & 93.6 & 93.4 & 93.4 &   - &   - & 92.8 & 93.3 & 93.3 &   - &   - & 87.2 & 86.8 & 86.9 &   - &   - \\ 
			\hline
			0 & \%zero &  99.9 & 100.0 & 100.0 & 100.0 & 100.0 & 100.0 & 100.0 & 100.0 & 100.0 & 100.0 & 100.0 & 100.0 & 100.0 & 100.0 & 100.0 \\ 
			& Bias & 0.04 & 0.00 & 0.00 & 0.00 & 0.00 & 0.00 & 0.00 & 0.00 & 0.00 & 0.00 & 0.00 & 0.00 & 0.00 & 0.00 & 0.00 \\ 
			& MSE & 0.00 & 0.00 & 0.00 & 0.00 & 0.00 & 0.00 & 0.00 & 0.00 & 0.00 & 0.00 & 0.00 & 0.00 & 0.00 & 0.00 & 0.00  \\ 
			\hline
	\end{tabular}}
\end{table}

\begin{table}[!p]
	\tblcaption{Performance of $\bbetahat\subDAC\ (\Isc=1,2,3)$ and $\bbetahat\subfulllin$ for estimating $\bbeta_0=\bbeta_0^{\sf(IV)}$ with respect to average computation time in seconds, GMSE ($\times 10^{-5}$), coefficient-specific empirical probability (\%) of $j\not\in\widehat\Asc$, bias ($\times 10^{-4}$), MSE ($\times 10^{-5}$), and empirical coverage probability (\%) of the confidence intervals.
		\label{Tbeta4}}
	{\tabcolsep=1.5pt
		\begin{tabular}{@{}cccccccccccccc@{}}
			\hline
			&  & \multicolumn{4}{c}{$v = 0.2$} & \multicolumn{4}{c}{$v = 0.5$} & \multicolumn{4}{c}{$v = 0.8$}  \\
			&  & \multicolumn{3}{c}{$\bbetahat\subDAC$} & $\bbetahat\subfulllin$
			& \multicolumn{3}{c}{$\bbetahat\subDAC$}
			& $\bbetahat\subfulllin$
			& \multicolumn{3}{c}{$\bbetahat\subDAC$}& $\bbetahat\subfulllin$\\
			& $\Isc=$  & $1$ & $2$ & $3$ &  & $1$ & $2$ & $3$ &  & $1$ & $2$ & $3$ &  \\ 
			\hline
			Time & &  62.0 & 120.8 & 178.8 & 262.1 &  58.0 & 112.3 & 166.4 & 263.5 &  58.3 & 113.8 & 168.6 & 253.9 \\ 
			GMSE & & 3.55 & 3.55 & 3.55 & 3.55 & 3.68 & 3.68 & 3.68 & 3.69 & 4.44 & 4.44 & 4.44 & 4.44 \\  
			\hline
			$p\subind=50$\\
			\hline
			0.08 & \%zero & 0.0 & 0.0 & 0.0 & 0.0 & 0.0 & 0.0 & 0.0 & 0.0 & 0.0 & 0.0 & 0.0 & 0.0 \\ 
			& Bias & 1.45 & 1.46 & 1.46 & 1.46 & 4.62 & 4.63 & 4.63 & 4.65 & 11.91 & 11.93 & 11.93 & 12.04 \\ 
			& MSE & 0.22 & 0.22 & 0.22 & 0.22 & 0.36 & 0.36 & 0.36 & 0.36 & 1.02 & 1.02 & 1.02 & 1.03 \\ 
			& CovP & 95.7 & 95.7 & 95.7 & 95.8 & 94.6 & 94.6 & 94.6 & 94.7 & 93.9 & 93.9 & 93.9 & 93.5 \\ 
			\hline
			0.04 & \%zero & 0.0 & 0.0 & 0.0 & 0.0 & 0.0 & 0.0 & 0.0 & 0.0 & 0.0 & 0.0 & 0.0 & 0.0 \\ 
			& Bias & -0.19 & -0.19 & -0.19 & -0.14 & 0.53 & 0.54 & 0.54 & 0.61 & 3.45 & 3.47 & 3.47 & 3.45 \\  
			& MSE & 0.22 & 0.22 & 0.22 & 0.22 & 0.33 & 0.33 & 0.33 & 0.33 & 0.93 & 0.93 & 0.93 & 0.93 \\ 
			& CovP & 95.4 & 95.5 & 95.5 & 95.5 & 96.0 & 96.1 & 96.1 & 96.0 & 93.7 & 93.8 & 93.8 & 93.7 \\ 
			\hline
			0.02 & \%zero & 0.0 & 0.0 & 0.0 & 0.0 & 0.0 & 0.0 & 0.0 & 0.0 & 0.0 & 0.0 & 0.0 & 0.0 \\ 
			& Bias &-4.55 & -4.56 & -4.56 & -4.58 & -6.79 & -6.78 & -6.78 & -6.89 & -17.43 & -17.42 & -17.42 & -17.21 \\ 
			& MSE & 0.25 & 0.25 & 0.25 & 0.25 & 0.45 & 0.45 & 0.45 & 0.45 & 1.59 & 1.59 & 1.59 & 1.57 \\ 
			& CovP & 94.1 & 94.1 & 94.1 & 94.4 & 90.5 & 90.8 & 90.8 & 90.8 & 86.8 & 86.7 & 86.7 & 86.8 \\ 
			\hline
			0 & \%zero & 100.0 & 100.0 & 100.0 & 100.0 & 100.0 & 100.0 & 100.0 & 100.0 & 100.0 & 100.0 & 100.0 & 100.0 \\ 
			& Bias & 0.00 & 0.00 & 0.00 & 0.00 & 0.00 & 0.00 & 0.00 & 0.00 & 0.00 & 0.00 & 0.00 & 0.00 \\  
			& MSE & 0.00 & 0.00 & 0.00 & 0.00 & 0.00 & 0.00 & 0.00 & 0.00 & 0.00 & 0.00 & 0.00 & 0.00 \\ 
			\hline
			$p\subdep=50$\\
			\hline
			0.08 & \%zero & 0.0 & 0.0 & 0.0 & 0.0 & 0.0 & 0.0 & 0.0 & 0.0 & 0.0 & 0.0 & 0.0 & 0.0 \\ 
			& Bias & 1.01 & 1.01 & 1.01 & 1.04 & 2.65 & 2.65 & 2.65 & 2.61 & 9.72 & 9.74 & 9.74 & 9.74 \\ 
			& MSE & 0.21 & 0.21 & 0.21 & 0.21 & 0.37 & 0.37 & 0.37 & 0.38 & 0.93 & 0.93 & 0.93 & 0.95 \\ 
			& CovP & 96.3 & 96.3 & 96.3 & 96.3 & 94.6 & 94.6 & 94.6 & 94.4 & 95.6 & 95.5 & 95.5 & 95.4 \\ 
			\hline
			0.04 & \%zero & 0.0 & 0.0 & 0.0 & 0.0 & 0.0 & 0.0 & 0.0 & 0.0 & 0.0 & 0.0 & 0.0 & 0.0 \\ 
			& Bias & -1.01 & -1.01 & -1.01 & -1.01 & 1.11 & 1.11 & 1.11 & 1.15 & 3.26 & 3.26 & 3.26 & 3.19 \\  
			& MSE & 0.21 & 0.21 & 0.21 & 0.21 & 0.33 & 0.33 & 0.33 & 0.33 & 0.93 & 0.93 & 0.93 & 0.94 \\
			& CovP & 95.8 & 95.9 & 95.9 & 96.0 & 95.2 & 95.2 & 95.2 & 95.3 & 94.9 & 94.8 & 94.8 & 95.0 \\ 
			\hline
			0.02 & \%zero & 0.0 & 0.0 & 0.0 & 0.0 & 0.0 & 0.0 & 0.0 & 0.0 & 0.0 & 0.0 & 0.0 & 0.0 \\ 
			& Bias & -3.81 & -3.81 & -3.81 & -3.81 & -6.28 & -6.26 & -6.26 & -6.30 & -16.47 & -16.47 & -16.47 & -16.59 \\
			& MSE & 0.26 & 0.26 & 0.26 & 0.25 & 0.45 & 0.45 & 0.45 & 0.45 & 1.49 & 1.48 & 1.48 & 1.49 \\ 
			& CovP & 93.8 & 93.8 & 93.8 & 93.5 & 91.2 & 91.1 & 91.1 & 91.4 & 86.2 & 86.1 & 86.1 & 86.0 \\ 
			\hline
			0 & \%zero & 100.0 & 100.0 & 100.0 & 100.0 & 100.0 & 100.0 & 100.0 & 100.0 & 100.0 & 100.0 & 100.0 & 100.0 \\ 
			& Bias & 0.00 & 0.00 & 0.00 & 0.00 & 0.00 & 0.00 & 0.00 & 0.00 & 0.00 & 0.00 & 0.00 & 0.00 \\ 
			& MSE & 0.00 & 0.00 & 0.00 & 0.00 & 0.00 & 0.00 & 0.00 & 0.00 & 0.00 & 0.00 & 0.00 & 0.00 \\ 
			\hline
	\end{tabular}}
\end{table}

% Figure 1 readmission within 30 days

% Figure 2 readmission due to PM within 30 days

\end{document}